\documentclass[letterpaper,titlepage,12pt]{article}
\usepackage[colorlinks=true,urlcolor=blue,citecolor=magenta]{hyperref}
\usepackage{amssymb,amsmath,amsfonts}
\usepackage{epsfig}
\usepackage{epsfig}
\usepackage{graphicx}
\usepackage{epstopdf}
\usepackage{caption}
\usepackage{subcaption}
\usepackage{amscd}
\usepackage{amsthm}
\usepackage{latexsym}
\usepackage{amsbsy}
\usepackage{bbm}
\usepackage[english]{babel}
\usepackage{psfrag}
\usepackage{tabularx}

\allowdisplaybreaks

\newcommand{\be}{\begin{eqnarray}}
\newcommand{\ee}{\end{eqnarray}}
\newcommand{\beq}{\begin{eqnarray}}
\newcommand{\eeq}{\end{eqnarray}}

\def\clock{{\count0=\time
           \divide\count0 60
           \ifnum\count0<10 0\fi\the\count0
           \multiply\count0 -60 \advance\count0 \time
           :\ifnum\count0<10 0\fi \the\count0
         }}
\newcommand{\timestamp}{{\small\vbox{\hbox{\tt\jobname.tex}
\hbox{\the\day/\the\month/\the\year, \clock}}}}

\numberwithin{equation}{section}

\begin{document}

\begin{titlepage}
\rightline{\vbox{   \phantom{ghost} }}
%

 \vskip 1.4 cm
\centerline{\LARGE \bf Gravitational Tension, Spacetime Pressure}
\vspace{.4cm}
\centerline{\LARGE \bf and Black Hole Volume}

\vskip 1.5cm

\centerline{\large {{\bf Jay Armas$^1$, Niels A. Obers$^2$ and Marco Sanchioni$^2$}}}

\vskip .8cm

\begin{center}

\sl $^1$ Physique Th\'eorique et Math\'ematique and International Solvay Institutes,\\
Universit\'e Libre de Bruxelles, C.P. 231, 1050 Brussels, Belgium.\\
\vspace{.3cm}
\sl $^2$ The Niels Bohr Institute, Copenhagen University,\\
\sl  Blegdamsvej 17, DK-2100 Copenhagen \O , Denmark.
\vskip 0.4cm

\end{center}
\vskip 0.6cm

\centerline{\small\tt  jarmas@ulb.ac.be, obers@nbi.ku.dk, sanchioni@nbi.ku.dk}

\vskip .8cm \centerline{\bf Abstract} \vskip 0.2cm \noindent
We study the first law of black hole thermodynamics in the presence of surrounding gravitational fields and argue that variations of these fields are naturally incorporated in the first law by defining gravitational tension or gravitational binding energy. We demonstrate that this notion can also be applied in Anti-de Sitter spacetime, in which the surrounding gravitational field is sourced by a cosmological fluid, therefore showing that spacetime volume and gravitational tension encode the same physics as spacetime pressure and black hole volume. We furthermore show that it is possible to introduce a definition of spacetime pressure and black hole volume for any spacetime with characteristic length scales which does not necessarily require a cosmological constant sourcing Einstein equations. However, we show that black hole volume is non-universal in the flat spacetime limit, questioning its significance. We illustrate these ideas by studying the resulting black hole volume of Kaluza-Klein black holes and of a toy model for a black hole binary system in five spacetime dimensions (the black saturn solution) as well as of several novel perturbative black hole solutions. These include the higher-dimensional Kerr-Newman solution in Anti-de Sitter spacetime as well as other black holes in plane wave and Lifshitz spacetimes.

\end{titlepage}

\pagestyle{empty}
\small
\tableofcontents

\normalsize
\newpage
\pagestyle{plain}
\setcounter{page}{1}


\newpage

\section{Introduction} \label{sec:intro}
The discovery that black holes carry an entropy \cite{Bekenstein:1973ur,Hawking:1974sw} proportional to the area of the event horizon has lead to view black holes as thermodynamic objects. This point of view provides a simple set of quantities, such as the mass $M$, the entropy $S$ and the angular momenta $J_{a}$, which can be used to characterise many of the properties of black holes. For an uncharged black hole in asymptotically flat spacetime, these quantities satisfy the first law of thermodynamics
\beq \label{eq:1stlaw}
dM=TdS+\sum_{a}\Omega_a dJ_a~~,
\eeq
where $T$ is the Hawking temperature and $\Omega_a$ is the set of horizon angular velocities. The study of \eqref{eq:1stlaw} leads to a deeper understanding of the dynamics, stability and uniqueness of these objects. Therefore, from a purely gravitational point of view it is important to understand what types and kinds of physical modifications can occur in \eqref{eq:1stlaw}. 

Many of such modifications are known and they arise due to other \emph{intrinsic} properties that black holes can have, such as an electric/magnetic charge or scalar hair \cite{Gibbons:1996af}. Other intrinsic properties such as horizon topology can allow for other types of charges, as in the case of the five-dimensional charged black ring \cite{Emparan:2004wy,Copsey:2005se}, which can cary dipole charge.\footnote{For other non-trivial topologies, generalisations of dipole charge were found in \cite{Emparan:2011hg}.} However, there are also \emph{extrinsic} properties that can affect \eqref{eq:1stlaw} such as the length scales characterizing the asymptotic region of a given black hole, or alternatively, the curvature scales characterising the resulting spacetime once the black hole is removed (i.e. when its horizon radius is set to zero).\footnote{Another external factor that can affect \eqref{eq:1stlaw} is non-trivial spacetime topology when, for example, there are fluxes present in the spacetime \cite{Gibbons:2013tqa}.}

A concrete physical set up which we have in mind, with potential astrophysical implications, is to understand what the modification of \eqref{eq:1stlaw} is when a black hole is immersed in the gravitational field of another black hole, as in a black hole binary system. In this case, each of the black hole horizons will satisfy a first law of thermodynamics of the form \eqref{eq:1stlaw} in which the mass (or horizon radius) of the other black hole appears as an external parameter.\footnote{Below we derive modified first laws for each individual horizon in a black hole binary system in a certain perturbative regime where one black hole can be seen as a probe in the background of the other. In Sec.~\ref{sec:bh} we show that this holds, beyond the leading order in this perturbative regime, for the black saturn solution \cite{Elvang:2007rd}, where the black ring can be seen as a probe in the background of the centre Myers-Perry black hole. While we are of the opinion that such individual first laws can be assigned to each individual horizon irrespective of such perturbative regime, further work is required in order to verify it.} A natural question to ask is then: how is the first law \eqref{eq:1stlaw} modified when there are variations in the gravitational field, in which the black hole is immersed, due to variations in the mass (or size) of the other black hole?

This question would be easily addressed if there existed exact analytic solutions of black hole binary systems. In four dimensional asymptotically flat spacetime no such solutions are known, however, perturbative solutions where a small black hole is orbiting a large black hole do exist up to several high orders \cite{Poisson:2009qj}. Unfortunately, these \emph{distorted} black hole solutions have only been constructed near the horizon and hence are not suitable for studying thermodynamic properties. Luckily, there are several exact or approximate analytic solutions that can be used as toy models for this kind of physics. This includes the first-order corrected solution \cite{Harmark:2003yz} for localized Kaluza-Klein black holes \cite{Harmark:2002tr}, which due to the periodicity of one of the coordinates and by the method of images, can be viewed as being immersed in their own gravitational field. Moreover, in five spacetime dimensions there are several examples of exact and analytic black hole binary systems in asymptotically flat spacetime. The simplest of these being the black saturn solution \cite{Elvang:2007rd}, in which a black ring horizon orbits the centre Myers-Perry black hole.\footnote{Other examples are bi-rings, di-rings and several multiple combinations of these \cite{Elvang:2007hs, Izumi:2007qx, Evslin:2007fv, Iguchi:2007is}.} Furthermore, the blackfold construction \cite{Emparan:2009cs, Emparan:2009at} provides a general tool to analytically construct perturbative solutions of large classes of black holes in non-trivial backgrounds, such as those considered in \cite{Caldarelli:2008pz, Camps:2008hb} along with the novel solutions with non-trivial spacetime asymptotics constructed in this paper. Indeed, we will use these examples to study the modifications to \eqref{eq:1stlaw}.

When studying the modifications to \eqref{eq:1stlaw} due to the presence of external gravitational fields, one wishes to introduce/observe new quantities which: \textbf{(1)} have a geometric/physical meaning, \textbf{(2)} have a thermodynamic interpretation, \textbf{(3)} can be defined in the presence of any gravitational field and \textbf{(4)} reduce to the same universal result once the gravitational field is removed. We wish to qualify these statements. 

By property \textbf{(1)} we mean that the quantities appearing in \eqref{eq:1stlaw} can be obtained, for a given black hole spacetime, e.g. by some integration over the horizon involving Killing vector fields (e.g. the Komar mass) or by looking at the asymptotic fall-off of the metric fields (such as the ADM mass). By property \textbf{(2)} we mean that all such quantities are clear analogues of classical thermodynamic quantities as the entropy $S$ or temperature $T$ and can be obtained by taking appropriate derivatives of the free energy. By property \textbf{(3)} we mean that such quantities can be defined for all black holes immersed in any gravitational field, regardless of what the source of that field might be. In fact, we seek to introduce a set of thermodynamic quantities which can be universally defined, regardless of the field being created by another black hole, by a star or by some cosmological fluid. Finally, by property \textbf{(4)} we mean that such quantities must have a universal limit when the gravitational field that surrounds the black hole is removed. Intuitively, one might think that any such extra quantity appearing in \eqref{eq:1stlaw} must vanish when the gravitational field is removed since in that case there are no other quantities characterising the (uncharged) black hole. However, for reasons that will become apparent later, we will not require from the start that such quantities must vanish in that limit but we will ultimately argue that the correct physical picture is one where such quantities do vanish in that limit.  

\subsection{Pressure and volume for Anti-de Sitter black holes}
As we have mentioned, we look for modifications of \eqref{eq:1stlaw} regardless of what is the source of the gravitational field. One can think of Anti-de Sitter (AdS) black holes in $D$ spacetime dimensions as being immersed in the gravitational field created by a cosmological fluid with pressure
\beq \label{pressure} 
P_e = -\frac{\Lambda }{8\pi G}~~,~~\Lambda=-\frac{(D-1)(D-2)}{2L^{2}}~~,
\eeq
which sources Einstein equations. In this case, variations in the gravitational field are controlled by variations of the cosmological constant $\Lambda$, or alternatively, by variations of the AdS radius $L$. In order to analyse the modifications of \eqref{eq:1stlaw} we consider the simplest case of the Schwarzschild-AdS black hole in $D=4$ with mass, entropy and temperature given by
\beq
M=\frac{r_+}{2G}\left(1+\frac{r_+^2}{L^2}\right)~~,~~S=\frac{\pi}{G} r_+^2~~,~~T=\frac{1}{4\pi r_+}\left(1+3\frac{r_+^2}{L^2}\right)~~,
\eeq
where $r_+$ is the horizon radius. Allowing for variations of the length scale $L$, one can easily verify that these black holes satisfy the first law of thermodynamics
\beq \label{eq:1stads}
dM=TdS+B_k d\mathbb{L}_k~~,
\eeq
where $\mathbb{L}_k$ is proportional to some power of $L$ such that $\mathbb{L}_k=\lambda L^{k}$, where $\lambda$ is an arbitrary constant which can depend on Newton's constant $G$ but otherwise cannot depend on any of the thermodynamic variables of the solution such as temperature $T$.  In turn, the response $B_k$ is given by
\beq \label{eq:master}
B_k=\left(\frac{\partial \mathcal F}{\partial \mathbb{L}_k}\right)_{T}~~,
\eeq
where $\mathcal F$ is the Gibbs free energy $\mathcal F=M-TS$. Moreover, this leads to a corresponding Smarr relation which follows from the Euler scaling argument
\beq \label{eq:smarr1}
(D-3)M-(D-2)TS=k B_k \mathbb{L}_k~~.
\eeq
From this point of view, any particular choice of $k$, from the infinite set of quantities $B_k$ and their infinite set of conjugate variables $\mathbb{L}_k$, is as good as any other in describing variations of the external gravitational field - a consideration which has not been previously stated in the literature.

A very popular choice in describing these variations in AdS has been the choice $k=-2$ and the identification $\mathbb{L}_{-2}$ of the extrinsic spacetime pressure, i.e. $\mathbb{L}_{-2}=P_e$ \cite{Kastor:2009wy,Dolan:2010ha,Cvetic:2010jb,Dolan:2011xt,Kubiznak:2012wp,Dolan:2013ft, El-Menoufi:2013pza}.\footnote{There is a large literature on considering the cosmological constant as a thermodynamic variable, starting with the early papers  \cite{Henneaux:1984ji,Teitelboim:1985dp, Henneaux:1989zc} and also for example the later work \cite{Caldarelli:1999xj,Sekiwa:2006qj,Wang:2006eb,Wang:2006bn,LarranagaRubio:2007ut,Urano:2009xn}.} In this case, $\mathbb{L}_{-2}$ has dimensions of pressure and the quantity $B_{-2}=V_i$ has dimensions of volume. For the particular case of the Schwarzschild-AdS black hole in $D=4$, this intrinsic black hole volume takes the form
\beq \label{eq:adsvol}
V_i=\frac{4}{3}\pi r_+^3~~.
\eeq
There are several interesting aspects of this particular choice. First of all, the quantities $P_e$ and $V_i$ have direct analogues with classical thermodynamic systems, they have dimensions of pressure and volume, respectively \cite{Dolan:2013dga, Dolan:2012jh, Dolan:2011jm}. In particular the phase diagram $P_e(V_i)$ for Schwarzschild-AdS black holes can be recast as a van der Waals equation \cite{Kubiznak:2012wp}\footnote{For earlier work in the same spirit see \cite{Chamblin:1999hg,Caldarelli:1999xj}.}
\beq \label{eq:pads}
P_e=\frac{T}{v}-\frac{1}{2\pi v^2}~~,~~v=2\left(\frac{3V_i}{4\pi}\right)^{\frac{1}{3}}~~,
\eeq
strengthening further the analogy with classical thermodynamic systems. Furthermore, the quantity $P_e$ is physically meaningful, since it is the pressure of the cosmological fluid, while $V_i$ has a geometric interpretation, as it can be obtained for a given black hole in AdS by means of evaluating the Killing potential \cite{Kastor:2009wy}. Moreover, there is something peculiar to the volume $V_i$. As it may be seen from \eqref{eq:adsvol}, the volume remains constant as the cosmological constant is sent to zero. The fact that this is the case and that the volume \eqref{eq:adsvol} coincides with the naive volume\footnote{By the \emph{naive volume} we mean the volume that can be obtained by taking the metric and performing a volume integration up to the horizon radius. For other black holes like rotating black holes there have been proposals for how to define the volume \cite{Parikh:2005qs,Ballik:2010rx,Ballik:2013uia} and we refer the reader to the review \cite{Dolan:2014jva} for a more detailed explanation of these cases.} in the flat spacetime limit has been seen as further strengthening the case for introducing the pair of thermodynamic variables $(P_e,V_i)$, despite the fact that no such quantity enters the first law \eqref{eq:1stlaw} for asymptotically flat black holes.

\subsection{A different point of view: gravitational tension}
We would like to understand whether or not the introduction of the set of variables $(P_e,V_i)$ is unique to AdS or can actually satisfy the desired properties \textbf{(1)}-\textbf{(4)} which we have described above. Regarding property \textbf{(2)} it is clear that it is satisfied by this set of variables, which also has property \textbf{(3)} since the exercise that we have performed for the Schwarzschild-AdS black hole can be carried out for any black hole with non-trivial asymptotics and because extra terms in the first law of the form \eqref{eq:1stads} can be obtained by performing the Legendre transform
\beq
M\to M+B_k\mathbb{L}_k~~,
\eeq
therefore moving onto extended phase space in which the length scale $\mathbb{L}_k$ is allowed to vary.

However, property \textbf{(1)} is not satisfied in a straightforward way. By this we mean that one cannot use Killing potentials to obtain the volume $V_i$ for any black hole in an arbitrary gravitational field, since the volume obtained by a suitable integral of the Killing potential is non-zero if there is a non-zero cosmological constant \cite{Kastor:2009wy}. As we will show in the course of this work, defining $V_i$ does not require a cosmological constant nor introducing matter in Einstein equations. In fact, it can be defined in backgrounds with non-trivial length scales which are solutions of the vacuum Einstein equations, such as plane wave solutions or black hole solutions. Moreover, the pair of variables $(P_e,V_i)$ does not satisfy property \textbf{(4)}, since as we will show in this paper, the volume $V_i$ in flat spacetime, obtained via the limit in which the background length scales are removed, is meaningless. We show this by taking two different charged black hole solutions both of which, when the length scales are removed, reduce to the same charged rotating asymptotically flat black hole but lead to two different volumes. In addition, one may study asymptotic plane wave black holes with an arbitrary number of length scales $L_a$ which, when $L_a=0$, lead to an arbitrary number of volumes describing an asymptotically flat black hole.\footnote{The idea of associating different pressures and conjugate volumes to different fields has been explored in \cite{Azreg-Ainou:2014lua, Azreg-Ainou:2014twa}. However, the approach of these works requires that the Einstein equations are sourced by extra fields, while here we take a more general point of view which does not require any extra source fields. Moroever, these works introduce further ad hoc assumptions such as that there is no pressure associated with components of the electromagnetic field. } In this sense, one would have to argue that an asymptotic flat black hole is characterized by an infinite set of volumes. 

These considerations demand another point of view and the introduction of new quantities which can be generally applied to any context where a gravitational field surrounding the black hole is present. We consider borrowing a concept which has its roots in the study of black branes and Kaluza-Klein (KK) black holes, namely, gravitational tension (or gravitational binding energy) \cite{Traschen:2001pb,Townsend:2001rg,Harmark:2004ch} and applying it in a broader context. 

Gravitational tension can be thought of as the contribution to the black hole energy due to the energy stored in the surrounding gravitational field. In the case of the black hole being a black brane, it is the same as the brane tension associated with a given non-compact direction. This notion of energy is described by the simplest choice of quantities in \eqref{eq:1stads}, namely, $k=1$ and $\lambda=1$ for which $\mathbb{L}_1=L$ and $B_1=\boldsymbol{\mathcal{T}}$. The first law of thermodynamics then takes the following form
\beq \label{eq:1stlaw2}
dM=TdS+\sum_{a}\Omega_a dJ_a+\Phi_{H}dQ_{(p)}+\sum_{a}\boldsymbol{\mathcal{T}}_{a}d L_a~~,
\eeq
where we have allowed for the presence of a $p$-form charge $Q_{(p)}$ and corresponding chemical potential $\Phi_H$ as well as the existence of several length scales $L_a$ and their corresponding tensions per unit length $\boldsymbol{\mathcal{T}}_{a}$. The corresponding Smarr relation \eqref{eq:smarr1} reads
\beq
\label{eq:smarr2}
(D-3)M-(D-2)\left(TS+\sum_{a}\Omega_{a}J_{ a}\right)-(D-3)\Phi_{\text{H}}Q= \hat{\boldsymbol{\mathcal{T}}} ~~,
\eeq
where $ \hat{\boldsymbol{\mathcal{T}}} $ is the total tension (or gravitational binding energy) given by
\beq \label{eq:tensionfree}
 \hat{\boldsymbol{\mathcal{T}}} =\sum_a \boldsymbol{\mathcal{T}}_{a} L_a=\sum_a L_a\left(\frac{\partial \mathcal F}{\partial L_a}\right)_{T,\Omega_a,\Phi_H}~~.
\eeq
From here we note that the total tension is obtained by summing the result of acting with the scaling operators $d/d\log L_a$ on the free energy. If we apply this to the case of the Schwarzschild-AdS black hole in $D=4$ for which $L_1=L$ and $\boldsymbol{\mathcal{T}}_1=\boldsymbol{\mathcal{T}}$ we obtain
\beq
\boldsymbol{\mathcal{T}}=-\frac{r_+^3}{L^3}~~.
\eeq
This quantity vanishes, as well as $ \hat{\boldsymbol{\mathcal{T}}}$, in the limit $L\to\infty$, in which the surrounding gravitational field is removed.

The introduction of these new pairs of variables $(L_a,\boldsymbol{\mathcal{T}}_{a})$ has several advantages. First of all, they have a well defined physical and thermodynamic meaning. If we take AdS spacetime as an example then $L_1=L$ is a measure of the \emph{spacetime volume} associated with each spacetime direction\footnote{Note that since we are using the word \emph{volume} associated to a given spacetime direction then this is equivalent to using the word \emph{length}.}, while $\boldsymbol{\mathcal{T}}_{a}$ is a measure of the energy stored per unit spacetime volume. In fact, when considering the case of black branes, $\boldsymbol{\mathcal{T}}_{a}$ is the brane tension per unit length, which is equal to minus the brane pressure. The variables $(L_a,\boldsymbol{\mathcal{T}}_{a})$ can be thought as the reverse of the variables $(P_e,V_i)$, in which brane tension has replaced spacetime pressure and spacetime volume has replaced black hole volume. Indeed, such point of view had already been taken in \cite{Harmark:2002tr, Harmark:2003dg,Kol:2003if, Harmark:2007md} for the particular case of KK black holes.

The variables $(L_a,\boldsymbol{\mathcal{T}}_{a})$ besides having a physical meaning, also have a well defined geometrical meaning. The length scales $L_a$ are simply the length scales associated with the curvature of spacetime along given spacetime directions. The gravitational tensions $\boldsymbol{\mathcal{T}}_{a}$, in turn, can be obtained in several different ways. If the spacetime has a compact direction, such as KK spacetimes, or if the the horizon is non-compact then one can simply apply the prescription of \cite{Harmark:2004ch}.\footnote{We note that this prescription does not require introducing Killing potentials.} If the black hole admits a blackfold limit, as large classes of higher-dimensional black holes do \cite{Emparan:2009vd, Armas:2010hz, Grignani:2010xm, Caldarelli:2010xz, Emparan:2011hg,Armas:2014bia, Armas:2015kra, Armas:2015nea}, then the prescription of \cite{Harmark:2004ch} also applies, once we zoom locally into the horizon rendering it brane-like. The total tension is subsequently obtained by integrating the local tension over the blackfold worldvolume. If the black hole does not admit such limit, as is the case of the Schwarzschild-AdS black hole, then the prescription of \cite{Harmark:2004ch} needs to be generalised and we leave this generalization for future work. For the moment, when dealing with these cases, we simply apply \eqref{eq:tensionfree}.

Finally, the set of variables $(L_a,\boldsymbol{\mathcal{T}}_{a})$ can be introduced in the presence of an arbitrary gravitational field and in this respect it is not different than the set of variables $(P_e,V_i)$. However, the variables $(L_a,\boldsymbol{\mathcal{T}}_{a})$ satisfy property \textbf{(4)} which the set $(P_e,V_i)$ does not. More precisely, when removing the gravitational field we find the universal result $\boldsymbol{\mathcal{T}}_{a}\to0$, which naturally does not lead to extra quantities describing an asymptotically flat black hole. 

\subsection{Brief summary}
In order to illustrate the ideas expressed above we will first consider the case of distorted black holes in Sec.~\ref{sec:bhc}, i.e., localized KK black holes which can be seen as black objects surrounded by the presence of their own gravitational field. In this context, we will show that the concept of tension is much more natural to introduce than the notion of black hole volume and we will already give evidence for the non-universality of black hole volume in the flat spacetime limit. 

In Sec.~\ref{sec:bf} we construct a series of new non-trivial and perturbative charged black hole solutions in Anti-de Sitter, plane wave and Lifshitz spacetimes using the blackfold approach. In here we study examples of spacetimes with multiple length scales, such as plane waves, and construct analogues of the higher-dimensional Kerr-Newman solution of \cite{Caldarelli:2010xz} in Anti-de Sitter and plane wave spacetimes. These new black hole solutions are interesting in their own right, in particular, the Kerr-Newman solution in AdS for which there is no corresponding exact solution. Furthermore, we provide the first example of a black hole which is rotating and moreover has non-trivial horizon topology in Lifshitz spacetimes. The reader may skip this section entirely if he/she is only concerned with the implications of these results to the modifications of \eqref{eq:1stlaw}.

In Sec.~\ref{sec:bh} we construct perturbative solutions in backgrounds with a black hole. One of these solutions corresponds to a specific limit of the black saturn solution. We then study in detail the example of this black hole binary system in five spacetime dimensions using the exact and analytic solution of \cite{Elvang:2007rd}. Here we focus on the case in which the Myers-Perry black hole in the centre is not rotating and show that the black ring horizon satisfies a first law of the form \eqref{eq:1stlaw2}.

Finally, in Sec.~\ref{sec:non}, we briefly use the new solutions of Sec.~\ref{sec:bf} to show that the notion of volume in flat spacetime is non-universal, while in Sec.~\ref{sec:out} we discuss some of the limitations of this work and future extensions of these research directions. We also provide some interesting results in the appendices, namely, in App.~\ref{app:thermo} we have collected all the thermodynamic properties of the new perturbative solutions constructed in this paper, while in App.~\ref{sec:odddipole} we have the explicit construction of a family of black holes carrying string charge. In App.~\ref{app:saturn} we give the thermodynamic quantities of the black saturn solution in the blackfold regime and compare it with our blackfold constructions.

\section{Black holes on cylinders: the Kaluza-Klein case }\label{sec:bhc}
In this section we analyse the thermodynamic properties of localized KK black holes which were constructed in a perturbative expansion in dimensions greater than four \cite{Harmark:2003yz}. These objects provide examples of black holes immersed in their own non-trivial gravitational field. We review that the thermodynamics of these objects follow \eqref{eq:1stlaw2}, as noted in \cite{Harmark:2003dg, Kol:2003if, Harmark:2003yz}, and that the tension can be extracted from the free energy using \eqref{eq:tensionfree}. We then show that the concept of black hole volume is not the desirable one when analysing variations in the surrounding gravitational field, which are controlled by variations in $L$ - the KK compactification parameter. In the end, we also take a look at the case of KK black strings.


\subsection{The localised black hole}
The localised black hole in KK spacetime is a static and perturbative solution found in \cite{Harmark:2003yz} in $D\ge5$, obtained by perturbing the Schwarzschild black hole to leading order in the parameter $r_0/L$ where $r_0$ is the horizon radius and $L$ the size of the KK circle. Its thermodynamic properties can be found in \cite{Harmark:2003yz} and read
\beq
M=\frac{\Omega_{(D-2)}}{16\pi G}(D-2)r_0^{D-3}\left(1+\frac{1}{2}\beta \left(\frac{r_0}{L}\right)^{D-3}\right)~~,
\eeq
\beq
T=\frac{(D-3)}{r_0}\left(1-\frac{(D-2)}{(D-3)}\beta \left(\frac{r_0}{L}\right)^{D-3}\right)~~,
\eeq
\beq
S=\frac{\Omega_{(D-2)}}{16\pi G}r_0^{D-2}\left(1+\frac{(D-2)}{(D-3)}\beta \left(\frac{r_0}{L}\right)^{D-3}\right)~~,
\eeq
where we have defined\footnote{Here $\zeta(s)$ is the Riemann Zeta function defined as $\zeta(s)=\sum_{m=1}^{\infty}m^{-s}$.}
\beq
\beta=\frac{\zeta(D-3)}{(2\pi)^{D-3}}~~.
\eeq
As noted already in \cite{Harmark:2002tr, Harmark:2003yz}, even though it is a localised black hole, it has a tension which can be obtained via the Smarr relation \eqref{eq:smarr2} and the above thermodynamic quantities. It reads
\beq
\hat{\boldsymbol{\mathcal{T}}}=\frac{\Omega_{(D-2)}}{32\pi G}\beta(D-2)(D-3)r_0^{D-3}\left(\frac{r_0}{L}\right)^{D-3}~~.
\eeq
We note that this quantity vanishes once we take the decompactification limit $L\to\infty$. This result can also be obtained using \eqref{eq:tensionfree} by evaluating the free energy
\beq
\mathcal F=\frac{\Omega_{(D-2)}}{16\pi G}r_0^{D-3}\left(1+\frac{1}{2}(D-2)\beta \left(\frac{r_0}{L}\right)^{D-3}\right)~~,
\eeq
and provides a non-trivial check of formula \eqref{eq:tensionfree}. The existence of this tension justifies the thermodynamic interpretation in terms of brane tension per unit length ${\boldsymbol{\mathcal{T}}}=\hat{\boldsymbol{\mathcal{T}}}/L$ and spacetime volume $L$. 

\subsubsection*{Defining spacetime pressure and black hole volume}
We now consider the possibility of defining a spacetime pressure and black hole volume. As we have noted in the previous section, black holes in spacetimes with non-trivial length scales satisfy \eqref{eq:1stlaw} for an infinite set of quantities. Following the same footsteps as in the introduction, we can attempt to define the black hole volume by choosing $k=-2$. This leads to the pressure $P_e=\lambda L^{-2} G^{-1}$ and the volume
\beq
V_i=-\frac{\Omega_{(D-2)}}{64\pi}\frac{L^2}{\lambda}\beta(D-2)(D-3)r_0^{D-3}\left(\frac{r_0}{L}\right)^{D-3}~~.
\eeq
Looking at the above expression, we notice that in dimensions $D>5$, the volume $V_i$ goes to zero in the decompactification limit $L\to\infty$, when one would expect that it would reduce to the non-zero volume of the Schwarzschild black hole in $D>5$ (see e.g. \cite{Altamirano:2014tva}) which was obtained by taking the flat spacetime limit of the volume of the Schwarzschild-AdS black hole,
\beq \label{eq:volschd}
V^{\text{sch}}_i=\frac{\Omega_{(D-2)}}{16\pi(D-1)}r_0^{D-1}~~.
\eeq
In $D=5$ we find that $V_i\propto r_0^{4}$ and hence we could choose $\lambda$ appropriately so that $V_i$ for KK black holes in $D=5$ would be equal to \eqref{eq:volschd} for the Schwarzschild black hole in the same number of spacetime dimensions. 

This definition of black hole volume, besides only making some sense in $D=5$, would also loose its geometric interpretation since in $D=5$ one would expect the black hole volume to be $V_i\propto r_0^{3}L$. It is possible to attempt defining the black hole volume by introducing a new length scale and defining a shifted mass. This new length scale $\tilde L$ leads to the right scaling of the volume in the decompactification limit $\tilde L\to\infty$ (or $L\to\infty$), namely,
\beq
\left(\frac{\tilde L}{r_0}\right)^2=\left(\frac{L}{r_0}\right)^{D-3}~~.
\eeq
In this way, we introduce the spacetime pressure $P_e=\lambda G^{-1}\tilde L^{-2}$ and define a new mass $\tilde M$ by shifting the mass $M$ by a fraction $b$ of the tension such that
\beq \label{eq:smass}
\tilde M=M+b\hat{\boldsymbol{\mathcal{T}}}~~.
\eeq
By requiring the correct thermodynamic behaviour, namely,
\beq
\frac{\partial \tilde M}{\partial S}|_{P_e}=T~~,
\eeq
this implies that we must choose
\beq
b=\frac{1}{D-3}-\frac{2}{D-1}~~.
\eeq 
Using \eqref{eq:master} with $k=-2$ and $\tilde L$ as the new length scale we obtain the black hole volume
\beq
V_i=-\frac{\Omega_{(D-2)}}{16\pi}\frac{\beta}{\lambda}\frac{(D-2)(D-3)}{(D-1)}r_0^{D-1}~~.
\eeq
Comparing this volume with the volume of the Schwarzschild black hole in $D$ dimensions \eqref{eq:volschd} we fix the factor $\lambda$ such that
\beq
\lambda=-\beta(D-2)(D-3)~~.
\eeq
While we see that introducing a new mass $\tilde M$, with a priori no inherent physical meaning, allows us to recover the Schwarzschild black hole volume, we believe that such possibility is not so natural. The concept of black hole volume and spacetime pressure is useful if the ADM mass $M$ would satisfy the first law \eqref{eq:1stlaw2}. However, this is not the case for this particular example. Consequently, we propose that the concept of tension $\hat{\boldsymbol{\mathcal{T}}}$ is a more useful one for studying variations in the external gravitational field for KK black holes.


\subsection{The black string}
Here we briefly consider the case of the KK black string. This solution is simply obtained by taking an asymptotically flat Schwarzschild black string and compactifying the infinitely extended direction on a circle of radius $L$. The resulting free energy for $D\ge5$ is given by
\beq
\mathcal{F}=\frac{\Omega_{(n+1)}}{16\pi G}r_0^{D-4} L~~,
\eeq
and once applying formula \eqref{eq:tensionfree} leads to the tension $\hat{\boldsymbol{\mathcal{T}}}=\mathcal{F}$. This tension naturally behaves like $\hat{\boldsymbol{\mathcal{T}}}\to\infty$ when we take the decompactification limit $L\to\infty$. This is expected since the total tension of an asymptotically flat black string diverges while the tension density ${\boldsymbol{\mathcal{T}}}=\hat{\boldsymbol{\mathcal{T}}}/L$ remains finite. This exercise had the purpose of showing that the tension per unit length of asymptotically flat black branes can be obtained by compactifying the infinitely extended directions on a circle and applying formula \eqref{eq:tensionfree}.

As in the previous case, we could introduce a naive definition of volume $V_i\propto r_0^{D-4}L^3$ or, by introducing a new length scale $\tilde L=r_0^2 L^{-1}$ and a new mass $\tilde M$ as in \eqref{eq:smass}, define a new volume of the form $V_i\propto r_0^{D-4} \tilde L$, which would give rise to \eqref{eq:volschd} in $D-1$ dimensions once we take the limit $\tilde L\to 0$ and evaluate the black hole volume per unit length $V_i/\tilde L$. The latter would require $b=-2/(D-2)$ and $\lambda=(D-4)/2$. However, for the same reasons as for the localized black hole, we find both of these possibilities unnatural.



\section{Blackfolds in background spacetimes with intrinsic length scales \label{sec:bf} } 
In this section we construct new perturbative (charged) black hole solutions in (Anti)-de Sitter, plane wave and Lifshitz spacetimes using the blackfold approach, which we first describe in Sec.~\ref{sec:essentials}. Of special importance is the perturbative construction of the Kerr-Newman solution in higher-dimensions both in (Anti)-de Sitter and plane wave spacetimes. Furthermore, we introduce the pair of variables $(\boldsymbol{\mathcal{T}},L_a)$ for all these solutions in order to describe variations in the surrounding gravitational field. This pertubative black hole solutions are interesting in their own right and provide evidence for the existence of a rich phase space of black hole solutions in these non-trivial spacetimes. If the reader is interested in the implications of these solutions to the notion of spacetime pressure and black hole volume, he/she can skip this section entirely and move on to Sec.~\ref{sec:bh}.


\subsection{Blackfold essentials} \label{sec:essentials}
In the blackfold approach \cite{Emparan:2009cs, Emparan:2009at}, one constructs stationary black objects with compact horizons by
starting with black brane solutions, with horizon scale $r_0$, of some (super)gravity theory and wrapping these on compact submanifolds with a characteristic scale $R$. In the regime $r_0 \ll R$ the near-horizon region is well approximated by a perturbed black brane geometry and can be corrected order-by-order in a derivative expansion \cite{Emparan:2007wm, Armas:2011uf, Camps:2012hw, Armas:2013hsa, Armas:2013goa, Armas:2014rva}. The prototypical case to keep in mind is the construction of an asymptotically flat black ring in any dimension $D \geq 5$ by wrapping a thin black string with horizon radius $r_0$ on a large circle with radius $R \gg r_0$ \cite{Emparan:2007wm}. 

This construction, however, is completely general and can be applied to the wrapping of black branes in any asymptotic background. Since it is convenient to work with known exact black brane solutions,
the explicit examples studied so far in the literature involve the bending/wrapping of asymptotically flat black branes. In this work, we are interested in bending asymptotically flat black branes carrying electric charge, which were found in \cite{Caldarelli:2010xz}.\footnote{We consider one example of branes carrying string charge in App.~\ref{sec:odddipole}.} These are black branes which are exact solutions of the following Einstein-Maxwell-dilaton action
\beq \label{eq:emd}
I=\frac{1}{16\pi G}\int d^{D}x\sqrt{-g}\left(R-2(\nabla\phi)^2-\frac{1}{4}e^{-2a\phi}F^{2}\right)~~,
\eeq
where $\phi$ is the dilaton field, with $a$ being the dilaton coupling and $F$ is the two-form field strength $dA$, with $A$ being the 1-form gauge field. It is convenient to introduce the parameter $N$ defined as
\beq
a^2=\frac{4}{N}-\frac{2(D-3)}{(D-2)}~~.
\eeq
The perturbative black hole solutions constructed from these brane geometries are not exclusively solutions of the action \eqref{eq:emd}. Instead, other terms, such as the cosmological constant and other field content, can be added to the above action without affecting this construction as long as the curvature scales associated with each new field are much larger than the horizon size $r_0$. More precisely, if the blackfold is being constructed in backgrounds with a set of intrinsic length scales $L_a$ then we must require that $r_0 \ll {\rm min} (R,L_a)$.\footnote{A more rigorous determination of the regime of validity of blackfold configurations consists in evaluating second order world volume scalar invariants \cite{Armas:2015kra}.} This implies that to leading order neither the curvature of the worldvolume nor the curvature associated with background scales are felt near the horizon and hence that locally the blackfold is still described by an asymptotically flat black brane solution of the action \eqref{eq:emd}. 

In order to locally wrap black branes, and in the absence of couplings to gauge/dilaton external fields, one must satisfy the local constraint equation, which can be derived from Einstein equations \cite{Emparan:2007wm, Camps:2012hw}, namely,
\beq \label{eq:bfext}
T^{ab}{K_{ab}}^{i}=0~~,
\eeq
where ${K_{ab}}^{i}$ is the extrinsic curvature tensor of the embedding geometry and $T^{ab}$ is the stress-energy tensor corresponding to the charged black brane, which in this case takes the perfect fluid form \cite{Caldarelli:2010xz}
\beq \label{eq:st}
\begin{split}
&T^{ab}=(\epsilon+P)u^{a}u^{b}+P\gamma^{ab}~~,~~(\epsilon+P)=-nP+\Phi \mathcal{Q}~~, \\
&P=-\frac{\Omega_{(n+1)}}{16\pi G}r_0^{n}~~,~~\Phi=\tanh\alpha~~,~~\mathcal{Q}=\frac{\Omega_{(n+1)}}{16\pi G}r_0^{n}n\sqrt{N}\sinh\alpha\cosh\alpha~~.
\end{split}
\eeq
Here we have introduced the fluid pressure $P$, energy density $\epsilon$, chemical potential $\Phi$, electric charge density $\mathcal{Q}$ and the induced metric on the $(p+1)$-dimensional geometry $\gamma_{ab}$ in dimensions $D=n+p+3$. The fluid variables $P,\epsilon, \mathcal{Q}$ depend only on the local temperature $\mathcal{T}$ and the local chemical potential $\Phi$, which are in turn functions of the coordinates $\sigma^{a}$ along the world volume. We have also introduced $\alpha$, which is the charge parameter of the brane and sometimes more convenient to parameterise blackfold solutions than the chemical potential $\Phi$.

In stationary equilibrium, which is the case we are interested here, as they give rise to stationary black holes, the fluid velocities $u^{a}$ must be aligned with a world volume Killing vector field $\textbf{k}^{a}$ with modulus $\textbf{k}$, which we can write, without loss of generality, as
\beq \label{eq:kvf}
\textbf{k}^{a}\partial_a=\partial_\tau + \sum_{\hat a}\Omega_{\hat a}\partial_{\phi_{\hat a}}~~,
\eeq
where $\tau$ is the time-like world volume direction and $\Omega_{\hat a}$ is the angular velocity associated with each of the Cartan angles $\phi_{\hat a}$.\footnote{The Killing vector field \eqref{eq:kvf} is required to map to a background Killing vector field \cite{Emparan:2009at}.} Furthermore, in equilibrium we also have that the global temperature $T$ and global chemical potential $\Phi_{\text{H}}$ are determined via a redshift of the local thermodynamic potentials such that $T=\textbf{k}\mathcal{T}$ and $\Phi_{\text{H}}=\textbf{k}\Phi$. This leads to the relation between the horizon size $r_0$ and the global thermodynamic potentials \cite{Caldarelli:2010xz},
\beq
r_0=\frac{n}{4\pi T}\textbf{k}\left(1-\frac{\Phi_{\text{H}}^2}{N\textbf{k}^2}\right)^{\frac{N}{2}}~~.
\eeq
A given stationary blackfold configuration is thus described by an induced world volume line element $\textbf{ds}^2=\gamma_{ab}d\sigma^{a}d\sigma^{b}$, a world volume Killing vector field $\textbf{k}^{a}$ and the global potentials $T$ and $\Phi_{\text{H}}$.

Stationary configurations may also have boundaries. In this case, the constraint equation \eqref{eq:bfext} must also be supplemented by the boundary condition
\beq
\textbf{k}|_{\partial\mathcal{W}_{p+1}}=0~~,
\eeq
which, in the uncharged case ($\Phi_{\text{H}}=0$) translates into the condition that the fluid must be moving at the speed of light on the boundary, while for non-zero charge, it has the physical interpretation that the brane must be extremal at the boundary. Stationary configurations have topologies of the form $\mathbb{R}\rtimes {\mathbb{B}}^{(p)}\rtimes \mathbb{S}^{(n+1)}$, where ${\mathbb{B}}^{(p)}$ is the topology of the spatial part ${\mathcal{B}}_p$ of the world volume geometry $\mathcal{W}_{p+1}$ and $\mathbb{S}^{(n+1)}$ denotes the topology of the transverse spherical space associated with the black brane geometry with properties \eqref{eq:st}.\footnote{Here we have assumed that the world volume geometry is of the form $\mathbb{R}\times {\mathcal{B}}_{p}$.} In the case of the existence of boundaries, ${\mathbb{B}}^{(p)}$ is not the topology of $\mathcal{B}_p$ but instead the result of a non-trivial fibration over $\mathcal{B}_p$.

In global thermodynamic equilibrium, the constraint equation \eqref{eq:bfext} can be equally derived from an effective free energy functional $\mathcal{F}$ given by \cite{Emparan:2011hg, Caldarelli:2010xz, Grignani:2010xm}
\beq\label{eq:free}
\mathcal{F}[X^{i}]=-\int_{\mathcal{B}_{p}}R_{0}P\left(\mathcal{T},\Phi\right)~~,
\eeq
where $X^{i}$ denotes the set of transverse scalars describing the position of the surface in the ambient space and $R_0$ is the modulus of the time-like worldvolume Killing vector field $\partial_\tau$. Given the effective free energy, one can easily extract the conjugate global thermodynamic potentials to $T, \Omega_{\hat a}, \Phi_{\text{H}}$ of the blackfold configuration, namely, the entropy $S$, the angular momenta $J_{\hat a}$ and the electric charge $Q$ via the corresponding relations
\beq\label{eq:thermo}
S=-\frac{\partial\mathcal{F}}{\partial T}~~,~~J_{\hat a}=-\frac{\partial\mathcal{F}}{\partial \Omega_{\hat a}}~~,~~Q=-\frac{\partial\mathcal{F}}{\partial \Phi_{\text{H}}}~~.
\eeq
These thermodynamic properties can also be obtained via the integration of appropriate conserved currents \cite{Emparan:2009at, Caldarelli:2010xz}. With these quantities introduced, we note that the free energy \eqref{eq:free} satisfies
\beq \label{eq:freerelation}
\mathcal{F}=M-TS-\sum_{a} \Omega_a J_a-\Phi_{\text{H}}Q_{(p)}~~.
\eeq
The total tension can also be obtained using \eqref{eq:tensionfree}, and using appropriate conserved currents, it is possible to derive its general form
\beq \label{eq:bftension}
\hat{\boldsymbol{\mathcal{T}}}
=-\int_{\mathcal{W}_{p+1}}dV_{(p)}R_0\left(\gamma^{ab}+n^{a}n^{b}\right)T_{ab}~~,~~n^{a}\partial_a=R_0^{-1}\partial_\tau~~.
\eeq
By the same token, it is possible to derive, from general principles, the Smarr relation \eqref{eq:smarr2} and the first law \eqref{eq:1stlaw2}. The Smarr relation \eqref{eq:smarr2} with the tension \eqref{eq:bftension} was first derived in \cite{Emparan:2009vd} and using the Euler argument, it follows that the first law \eqref{eq:1stlaw2} is satisfied. We thus see that blackfolds naturally exhibit this universal thermodynamic behavior in spacetimes with non-trivial asymptotics.


\subsection{(Anti)-de Sitter background}
Here we construct novel black holes with electric charge in global (A)dS and in App.~\ref{sec:odddipole} we consider the case of black holes with dipole charge. We write the global (A)dS metric in the form
\beq\label{eq:dsads}
ds^2=-f(r)dt^2+f(r)^{-1}dr^2+r^2d\Omega_{(D-2)}^2~~,~~f(r)=1+\frac{r^2}{L^2}~~,
\eeq
where $L$ is the AdS radius. In order to obtain the de Sitter metric we simply perform the Wick rotation $L\to i L$.


\subsubsection{Charged black odd-spheres} \label{sec:oddads}
This type of configurations are obtained by embedding a $p$-dimensional sphere with radius $R$ in the background \eqref{eq:dsads}. We set the configuration to rotate with equal angular velocity $\Omega$ in each of the $[(p+1)/2]$ Cartan angles of the $p$-dimensional sphere, labelled by $\phi_{\hat a}$. The embedded geometry and the corresponding Killing vector field are given by
\beq \label{eq:dsodd}
\textbf{ds}^2=-f(\textbf{R})d\tau^2+R^2d\Omega_{(p)}^2~~,~~\textbf{k}^{a}\partial_a=\partial_\tau+\Omega\sum_{\hat a=1}^{[(p+1)/2]}\partial_{\phi_{\hat a}}~,~f(\textbf{R})=1+\textbf{R}^2~,
\eeq
where we have defined the dimensionless radius $\textbf{R}=R/L$. We choose to parametrise the resulting configuration in terms of the variables $r_0,\textbf{R},\alpha$. In terms of these, the free energy takes the simple form
\beq \label{eq:freeodd}
\mathcal{F}[R]=\frac{\Omega_{(n+1)}V_{(p)}}{16\pi G}f(\textbf{R})r_0^{n}~~,
\eeq
where $V_{(p)}=\Omega_{(p)}R^{p}$. Upon variation of this with respect to $R$, restricting to the cases where $p$ is odd, and solving the resulting equation leads to the equilibrium condition
\be \label{eq:eqads}
\Omega^2 R^2=(1+\bold R^2)\frac{\bold R^2\left(n N \sinh ^2\alpha+n +p+1\right)+p}{ \bold R^2 \left(n N \sinh ^2\alpha+n +p+1 \right)+n N \sinh ^2\alpha+n +p}~~.
\ee
This expression connects to several others in the literature. In the uncharged limit $\alpha=0$, this yields the result obtained in \cite{Armas:2010hz}, while in the flat space limit $L\to\infty$, this yields the result  of \cite{Caldarelli:2010xz}. When taking both limits, $\alpha\to0$ and $L\to\infty$ we obtain the result of \cite{Emparan:2009vd}.

\subsubsection*{Properties of the solution}
There are several interesting cases that should be noted. First of all, these black holes with horizon topology $\mathbb{R}\times \mathbb{S}^{(p)}\times \mathbb{S}^{(n+1)}$ admit an extremal limit. Taking $\alpha\to\infty$ we obtain the equilibrium condition
\beq
\Omega^2 R^2=\frac{1+\bold R^2}{2}~~,
\eeq
for extremal black odd-spheres. In the flat space case for which $\bold{R}=0$ this reduces to the analysis of \cite{Caldarelli:2010xz}.

Secondly, in the deSitter case for which $L\to i L$ and hence $\bold{R}\to i\bold{R}$, not all values of $\bold{R}$ are allowed. In fact we find the two possible regimes
\be
\bold R\leq 1\quad \vee\quad \frac{p}{n N \sinh ^2\alpha+n +p+1}\leq \bold R\leq\frac{n N \sinh ^2\alpha+n +p}{n N \sinh ^2\alpha+n +p+1}~~.
\ee
The extremal branch of solutions lies within the regime $\bold R\leq 1$. Furthermore, as noted in \cite{Armas:2010hz}, in deSitter spacetime there can exist static solutions ($\Omega=0$) and, as noted in \cite{Armas:2015kra}, they are valid for all $p\ge1$ and not only for odd $p$. These are solutions for which the radius takes the specific value of
\be
\bold R= \frac{p}{n N \sinh ^2\alpha+n +p+1}~~.
\ee
We note here that this branch of static solutions does not admit an extremal limit. 

\subsubsection*{Gravitational tension}
Using \eqref{eq:freeodd} we can obtain all thermodynamic properties which we collect in App.~\ref{app:thermo} while here we present expressions for the tension. Using formula \eqref{eq:tensionfree} together with \eqref{eq:freeodd}, we obtain the total tension given by
\beq
\hat{\boldsymbol{\mathcal{T}}}=-\frac{ \Omega_{(n+1)}V_{(p)}}{16 \pi  G}r_0^n\bold R^2 \sqrt{1+\bold R^2} \left(nN\sinh^2\alpha+n+p+1\right)~~,
\eeq
which is a function of $T,\Omega$ and $L$, where $\Omega$ was given in \eqref{eq:eqads}. From here we can introduce the tension per unit spacetime volume such that
\beq
{\boldsymbol{\mathcal{T}}}=\frac{\hat{\boldsymbol{\mathcal{T}}}}{L}=-\frac{ \Omega_{(n+1)}V_{(p)}}{16 \pi  G}\frac{r_0^n}{L}\bold R^2 \sqrt{1+\bold R^2} \left(nN\sinh^2\alpha+n+p+1\right)~~.
\eeq
These quantities satisfy the first law \eqref{eq:1stlaw2} and the Smarr relation \eqref{eq:smarr2}. Furthermore, both these quantities vanish once we take the limit $L\to\infty$, leaving their corresponding asymptotically flat counterparts with no extra quantities characterizing them.


\subsubsection{Charged black discs: analogue of the Kerr-Newman black hole} \label{sec:discads}
In this section we make a perturbative construction of the analogue of the Kerr-Newman black hole in higher-dimensional (A)dS with one single angular momentum. This corresponds, in the blackfold approximation, to an electrically charged rotating disc with induced metric and Killing vector field
\beq
\textbf{ds}^2=-f(\rho)d\tau^2+f(\rho)^{-1}d\rho^2+\rho^2d\phi^2~,~\textbf{k}^{a}\partial_a=\partial_\tau+\Omega\partial_\phi~,~f(\rho)=1+\frac{\rho^2}{L^2}~.
\eeq
This geometry develops a boundary when $\textbf{k}=0$, for which the brane becomes locally extremal, corresponding to the maximum of $\rho$ given by
\beq \label{eq:discmax}
\rho_{\text{max}}=\frac{\sqrt{1-\Phi_{\text{H}}^2/N}}{\Omega\sqrt{\xi}}~~,~~\xi=1-\frac{1}{L^2\Omega^2}~~,
\eeq
which implies in the AdS case that $\Omega L\ge1$. This disc configuration trivially solves the blackfold equations \eqref{eq:bfext} since it is a minimal surface \cite{Armas:2015kra}. The thickness of the disc and the charge parameter are given by
\be \label{eq:r0disc}
\begin{split}
&r_0(\rho)=\frac{n}{4\pi T}\left(1-\xi  \rho^2 \Omega ^2\right)^{\frac{1-N}{2}} \left(1-\xi  \rho^2 \Omega ^2-\frac{\Phi_\text{H} ^2}{N}\right)^{N/2}~~,\\
&\tanh\alpha(\rho)=\frac{\Phi_\text{H}/\sqrt{N} }{ \sqrt{1-\xi  \rho^2 \Omega ^2}}~~.
\end{split}
\ee
Therefore, at the boundary $\rho_{\text{max}}$ the thickness $r_0$ of the disc vanishes and hence the resulting black holes have topology $\mathbb{R}\times \mathbb{S}^{(D-2)}$. The thickness remains finite for all values of $\xi$ and hence this configuration lies within the regime of validity $r_0\ll L$.\footnote{The boundary of the blackfold deserves special attention, see \cite{Armas:2015kra} for a discussion of its validity.} In the uncharged case $\Phi_\text{H}=0$, this reduces to the construction of \cite{Armas:2010hz} and when $L\to\infty$, hence when $\xi\to1$, it reduces to the higher-dimensional Kerr-Newman solution perturbatively constructed in \cite{Caldarelli:2010xz}.

This configuration has several interesting properties. In particular, it admits an extremal limit, for which $\Phi_{\text{H}}\to\sqrt{N}$. In this case it seems that the size of the disc \eqref{eq:discmax} would shrink to zero. As noted in \cite{Caldarelli:2010xz} for the asymptotically flat case, one must also have that $\Omega\to0$ such that the ratio $\sqrt{1-\Phi_{\text{H}}^2/N}/\Omega$ remains finite. However, in the presence of the cosmological constant this is not possible. Instead, attaining extremal regimes is only possible in AdS for which one must send $\Omega L\to 1$, hence $\xi\to0$, such that the ratio $\sqrt{1-\Phi_{\text{H}}^2/N}/\sqrt{\Omega^2L^2-1}$ remains constant. From \eqref{eq:r0disc} it implies that the temperature must also approach zero such that the ratio $(1-\Phi_{\text{H}}^2/N)^{N/2}/T$ remains finite. In this situation, the disc is not rotating at very slow speeds as in the flat space case. This is not possible in deSitter spacetime as there one has that $\xi\ge1$. On the other hand we note that in dS, static solutions where the disc has finite size exist for which its size is given by
\beq
\rho_{\text{max}}|_{\Omega\to0}=L\sqrt{1-\Phi_{\text{H}}^2/N}~~.
\eeq 
When the disc is uncharged and static, it ends on the cosmological horizon as noted in \cite{Armas:2010hz}. However, we observe here that if the disc is charged it does not reach the cosmological horizon. One may also conclude from here that static and extremal discs do not lie within the regime of validity of our method.

We now proceed and evaluate the free energy for these configurations from which all thermodynamic properties can be obtained. This is given by
\beq \label{eq:freedisc}
\mathcal{F}=\frac{\Omega_{(n+1)}}{8G}\tilde r_0^n\frac{\,_2F_1\left(1,\frac{1}{2} (N-1) n;\frac{N n}{2}+2;1-\frac{\Phi^2_{\text{H}}}{N}\right)}{\xi\Omega^2(2+N n) }~~,
\eeq
where we have defined
\beq
\tilde r_0^n=\left(\frac{n}{4\pi T}\right)^{n}\left(1-\frac{\Phi ^2_{\text{H}}}{N}\right)^{\frac{2+N n}{2}}~~.
\eeq
The thermodynamic properties are given in App.~\ref{app:thermo}. From here we also extract the tension
\be\label{eq:volads}
\begin{split}
\hat{\boldsymbol{\mathcal T}}&= -\frac{\Omega_{(n+1)}}{4G}\tilde r_0^n\frac{\,
   _2F_1\left(1,\frac{1}{2} (N-1) n;\frac{Nn}{2}+2;1-\frac{\Phi
   ^2_{\text{H}}}{N}\right)}{(2+N n) \xi^2L^2\Omega^4}~~,
 \end{split}
\ee
which, as expected, vanishes in the limit $L\to\infty$. Its thermodynamic properties satisfy \eqref{eq:1stlaw2} and \eqref{eq:smarr2}.


\subsection{Plane wave background} \label{sec:pp}
In this section we consider the perturbative construction of new black holes in a plane wave background with metric
\beq \label{eq:dspp}
ds^2=-(1+\sum_{q=1}^{D-2}A_q x_q^2)dt^2+(1-\sum_{q=1}^{D-2}A_q x_q^2)dy^2-2\sum_{q=1}^{D-2}A_q x_q^2dtdy+\sum_{q=1}^{D-2}dx_q^2~~.
\eeq
Here the constants $A_q$ define the curvature length scales $1/\sqrt{A_q}$ of the spacetime. We furthermore assume that this is a solution of the vacuum Einstein equations, which requires that $\text{Tr}A_q=0$. This background is of special importance, since contrary to (A)dS, it is in general anisotropic as there is a set of $(D-3)$ independent length scales, each of them associated with a particular spacetime direction. 

The configuration studied in (A)dS in Sec.~\ref{sec:oddads} is also a solution in the background \eqref{eq:dspp} as long as it is embedded such that $y=0$ \cite{Armas:2015kra}. To leading order in this perturbative construction all the solution properties are the same provided we define $\bold R =R \sqrt{A_1}$ where $A_1$ is the value of $A_q$ for all the directions involved in the planes of rotation of the odd-sphere. In order to exhibit the differences between this spacetime and (A)dS, we consider the case of product of odd-spheres, to highlight its multi-scale properties, and the analogue of the disc solution of Sec.~\ref{sec:discads}.


\subsubsection{Products of black odd-spheres: a multi scale example} \label{sec:planeodd}
In this section we consider the simple case of the product of $m$ uncharged odd-spheres in the background \eqref{eq:dspp}. The corresponding black holes solutions have horizon topology $\mathbb{R}\times\prod_{\hat a=1}^{n}\mathbb{S}^{(p_{\hat a})}\times \mathbb{S}^{(n+1)}$ and were in fact constructed in \cite{Armas:2015kra}. These are described by the induced metric and Killing vector field
\beq
\!\!\!\!\!\!\textbf{ds}^2=-R_0^2d\tau^2+\sum_{\hat a=1}^{m}R_{\hat a}^2d\Omega_{(p_{\hat a})}^2~,~R_0^2=1\!+\!\sum_{\hat a=1}^{m}A_{\hat a}R_{\hat a}^2~,~\textbf{k}^{a}\partial_a=\partial_\tau+\sum_{\hat a=1}^{[(p+1)/2]}\!\!\!\Omega^{\hat a}\partial_{\phi_{\hat a}}~,
\eeq
where $\Omega^{\hat a}$ is the angular velocity associated with all the $[(p_{\hat a}+1)/2]$ Cartan angles of each $p_{\hat a}$-dimensional sphere. It's free energy is given by
\beq \label{eq:freeoddpp}
\mathcal{F}[R_{\hat a}]=-V_{(p)}R_0 P~~,~~V_{(p)}=\prod_{\hat a=1}^{m}\Omega_{(p_{\hat a})}R_{\hat a}^{p_{\hat a}}~~,
\eeq
while the equilibrium condition reads \cite{Armas:2015kra}
\beq
(\Omega^{\hat a})^{2}R_{\hat a}^2=R_0^2\frac{\hat p_a+\bold R_{\hat a}^2(n+p+1)}{(n+p)+(n+p+1)\bold R^2}~~,~~\bold R_{\hat a}^2=A_{\hat a}R_{\hat a}^2~~,~~\bold R=\sum_{\hat a=1}^{m}\bold R_{\hat a}^2~~.
\eeq
The thermodynamic properties, namely, the mass, angular momenta and entropy can actually be found in \cite{Armas:2010hz} since these configurations to leading order exhibit the same properties as in (A)dS if one identifies $A_{\hat a}=L^{-2}$ for all $\hat a=1,..,m$. Note that here we assume that there is another non-zero scale $A_{m+1}$ such that $\text{Tr}A_q=0$ but which does not enter in the thermodynamics of the configuration due to the choice of embedding. Hence all the scales $A_{\hat a}$ with $\hat a=1,..,m$ are independent scales.

We now proceed and analyse the total tension $\hat{\boldsymbol{\mathcal{T}}}$ from \eqref{eq:freeoddpp}, which is given as a sum of tensions, one for each $p_{\hat a}$-sphere, which we label by $\hat{\boldsymbol{\mathcal{T}}}_{\hat a}$. We find the simple result\footnote{Note that here we had to use instead the formula $\hat{\boldsymbol{\mathcal{T}}}_{\hat a}=-2A_{\hat a}\frac{\partial \mathcal{F}}{\partial A_{\hat a}}$ due to the the fact that $A_{\hat a}$ has dimensions of inverse length square.}
\beq \label{eq:tensionpp}
\hat{\boldsymbol{\mathcal{T}}}_{\hat a}=-\frac{ \Omega_{(n+1)}V_{(p)}}{16 \pi  G}r_0^n\bold R_{\hat a}^2 \sqrt{1+\bold R^2} \left(n +p+1\right)~~,
\eeq
which vanishes when $A_{\hat a}\to0$. This example shows that these black holes are characterised by a set of tensions, satisfying the first law \eqref{eq:1stlaw2}.


\subsubsection{Charged black discs} \label{sec:discpp}
In this section we construct the analogue of the disc solution of Sec.~\ref{sec:discads}, which will reveal the non-universal character of the spacetime pressure for a given spacetime. These configurations are charged versions of those found in \cite{Armas:2015kra}, obtained via an embedding such that $y=0$ with induced metric and Killing vector field
\beq \label{eq:dsdiscpp}
\textbf{ds}^2=-R_0^2d\tau^2+d\rho^2+\rho^2d\phi^2~~,~~R_0^2=1+A_1\rho^2~~,~~\textbf{k}^{a}\partial_a=\partial_\tau+\Omega\partial_\phi~~,
\eeq
where we have chosen $A_2=A_1$ such that the Killing vector field presented above is a Killing vector field of the background \eqref{eq:dspp}. This is trivially a solution of the blackfold equations as it also represents a minimal surface in these spacetimes \cite{Armas:2015kra}. The thickness and charge parameter of the disc are given by \eqref{eq:r0disc} but with the replacement $A_1=L^{-2}$. Hence, the disc has a maximum size given by \eqref{eq:discmax} and the discussion regarding extremal limits in AdS also holds in this case provided $A_1>0$. The case $A_1<0$ is similar to deSitter spacetime. 

The thermodynamic properties, due to the different induced metric, are however altogether distinct from its (A)dS counterpart. It is possible to derive them for any $N$, however, the resulting expressions are slightly cumbersome. Therefore we focus on the simplest case of $N=1$ which captures all the essential physics.\footnote{The case $N=1$ corresponds to a Kaluza-Klein reduction.} In this case, the free energy takes the following form
\beq
\!\!\!\!\!\mathcal{F}=\frac{\Omega_{(n+1)}}{8 G}\tilde r_0^n\frac{ \,
   _2F_1\left(-\frac{1}{2},1;\frac{n+4}{2};\frac{A_1 (\Phi_{\text{H}}^2 -1) }{\Omega ^2-A_1}\right)}{(n+2) \left(\Omega ^2-A_1\right)}~~,~~\tilde r_0^n=\left(\frac{n}{4\pi T}\right)^{n}\left(1-\Phi ^2\right)^{\frac{n}{2}+1}~,
\eeq
while the remaining thermodynamic properties are given in App.~\ref{app:thermo}. From here we extract, as previously, the total tension and find
\beq \label{eq:tensiondiscpp}
\begin{split}
\!\!\!\!\!\!\hat{\boldsymbol{\mathcal{T}}}=&-\frac{\Omega_{(n+1)}}{2 G}\tilde r_0^n\frac{ \Gamma \left(\frac{n}{2}+3\right)}{(n+2) (n+4)
   \left(A_1-\Omega ^2\right)^2} \Big(\Omega ^2 \,
   _2\tilde{F}_1\left(-\frac{1}{2},2;\frac{n+4}{2};\frac{A_1 (\Phi^2_{\text{H}} -1)
   }{\Omega ^2-A_1}\right) \\
   &+\left(A_1-\Omega ^2\right) \,
   _2\tilde{F}_1\left(-\frac{1}{2},1;\frac{n+4}{2};\frac{A_1 (\Phi_{\text{H}}^2 -1)
   }{\Omega ^2-A_1}\right)\Big)~~,
\end{split}
\eeq
which vanishes, as expected, when $A_1\to0$.


\subsection{Lifshitz background} \label{sec:lifs}
We now consider one of the Lifshitz spacetimes found in \cite{Tarrio:2011de} as the background for perturbatively constructing new solutions. This spacetime has a spherically symmetric metric of the form
\beq
ds^2=-\frac{r^{2z}}{L^{2z}}f(r)dt^2+\frac{L^2}{r^2}f(r)^{-1}dr^2+r^2d\Omega_{(D-2)}^2~~,~~f(r)=1+\beta\frac{L^2}{r^2}~~,
\eeq
where we have defined the constant $\beta$ as
\beq \label{eq:lifbeta}
\beta=\left(\frac{(D-3)}{(D+z-4)}\right)^2~~.
\eeq
Here, $z$ is the Lifshitz exponent which can lie in the interval $z\ge1$. In the limit $z\to1$ we obtain the AdS metric \eqref{eq:dsads}. This spacetime is in fact supported by several gauge fields and dilaton in order to be a solution of the model studied in \cite{Tarrio:2011de}. However, since we will be constructing an uncharged solution, we do not need to take into account possible couplings to these background fields. 

We focus on the analogue configurations of the odd-spheres of Sec.~\ref{sec:oddads}, hence of a class of black holes with horizon topology $\mathbb{R}\times\mathbb{S}^{(p)}\times\mathbb{S}^{(n+1)}$ which includes the case of the black ring ($p=1$). This configuration has an induced metric and Killing vector field given by
\beq
\!\!\!\bold{ds}^2=-\frac{R^{2z}}{L^{2z}}f(R)d\tau^2+R^2d\Omega_{(p)}^2~,~f(R)=1+\beta\frac{L^2}{R^2}~,~\bold{k}^{a}\partial_a=\partial_\tau+\Omega\!\!\!\sum_{\hat a=1}^{[(p+1)/2]}\!\!\!\partial_{\phi_{\hat a}}~.
\eeq
The blackfold equations \eqref{eq:bfext} are easily solved. By defining $\bold R=R/L$ we obtain the equilibrium condition
\be
\Omega R=\frac{\sqrt{\beta +\bold R^2} \bold R^{z-1} \sqrt{(n+2) \left(\bold R^2 z+\beta  (z-1)\right)+p \left(\beta +\bold R^2\right)}}{\sqrt{\left(\beta +\bold R^2\right)~~.
   (n+p+2z-1)-\beta }}~~.
\ee
With this we evaluate the horizon size of the odd-sphere and find
\beq \label{eq:lifr0}
r_0=\frac{n}{4\pi T} \left(-\frac{n \left(\beta +\bold R^2\right) \bold R^{2 z-2} \left(\bold R^2 (z-1)+\beta  (z-2)\right)}{\left(\beta +\bold R^2\right)
   (n+p+2z-1)-\beta}\right)^{n/2}~~.
\eeq
We see that for this to take real values we must require that 
\beq \label{eq:lifcond}
\bold R^2 (z-1)+\beta (z-2)<0~~,
\eeq
which implies that we must restrict to values of $z$ within the interval $1\le z<2$. This we assume from now on and in the remaining of this section.

With this we obtain the total tension which takes the form
\be
\hat{\boldsymbol{\mathcal T}} =-\frac{\Omega_{(n+1)}}{16\pi G}V_{(p)}r_0^{n}\frac{(n+p+z) \sqrt{\beta +\bold R^2} \left(z \left(\beta + \bold R^2\right)-\beta \right) \bold R^{z-1} }{|\bold R^2 (z-1)+\beta  (z-2)|}~~,
\ee
which vanishes when $L\to\infty$. The thermodynamic quantities non-trivially satisfy the first law \eqref{eq:1stlaw2}. This is the first example of a black hole solution with non-trivial topology in these spacetimes.

\section{Blackfolds in background spacetimes with a black hole \label{sec:bh} }
In this section we present the perturbative construction of toy models of black hole binary systems, namely, the generalisation of the black saturn solution to higher-dimensions and with electric charge. Focusing on the case in which the centre black hole is not rotating and is surrounded by a black ring, leaving the rotating case for future work, we consider introducing a definition of black hole volume. We show that such definition does not give rise to the expected scaling once the centre black hole is removed. Therefore, using gravitational tension as the natural modification of \eqref{eq:1stlaw}, we take the exact black saturn solution in five dimensions and extract the tension to one higher-order in the ultraspinning regime. This requires enforcing the first law \eqref{eq:1stlaw2} to hold on the black ring horizon.

Finally, we also consider the generalisation of these solutions to AdS backgrounds, thereby illustrating the case in which there are two different tensions, where one is associated with the centre black hole mass and the other with the cosmological constant.

\subsection{Schwarzschild black hole background} \label{sec:sch}
We begin with the Schwarzschild black hole background which has a spherically symmetric metric as in \eqref{eq:dsads} but with a function $f(r)$ given by
\beq \label{eq:fsch}
f(r)=1-\frac{\mu^{D-3}}{r^{D-3}}~~,
\eeq
where $\mu$ is the horizon location of the Schwarzschild black hole. In this background we place a charged odd-sphere geometry at a fixed $r=R$ and hence the induced metric and Killing vector field are the same as in Sec.~\ref{sec:oddads} as well as the free energy \eqref{eq:freeodd} but with the function \eqref{eq:fsch}. It is straightforward to obtain the equilibrium condition, which reads
\be\label{eq:equigen}
\Omega^2R^2=f(R)\frac{2pf(R)+(1+n+nN\sinh^2\alpha) R f '(R)}{2(nN\sinh^2\alpha+n+p)f(R)+Rf'(R)}~~.
\ee
This equilibrium condition is in fact valid for any spherically symmetric spacetime of the form \eqref{eq:dsads} for some function $f(r)$. In the limit $\alpha\to0$ it reduces to that obtained in \cite{Armas:2010hz}. Defining the dimensionless ratio $\bold R= \mu/R$, which vanishes when the black hole is removed $\mu\to0$, and using \eqref{eq:fsch} we can rewrite the above condition as
\be \label{eq:solscz}
\Omega^2 R^2= \frac{ \left(\bold R^{n+p}-1\right) \left(\bold R^{n+p} \left(n N \sinh ^2\alpha  (n+p)+n (n+p+1)-p\right)+2 p\right)}{ \left(2 n N
   \sinh ^2\alpha  \left(\bold R^{n+p}-1\right)+(n+p) \left(\bold R^{n+p}-2\right)\right)}~~.
\ee
From here we see that extremal configurations exist when $\alpha\to\infty$. Furthermore, using the free energy \eqref{eq:freeodd} in order to compute the conserved charges of this configuration, which are presented in App.~\ref{app:thermo}, and the Smarr relation \eqref{eq:smarr2}, we obtain the total tension
\beq \label{eq:tensch}
\hat{\boldsymbol{\mathcal{T}}}=-\frac{\Omega_{(n+1)}V_{(p)}r_0^{n}}{16\pi G}\bold R^{n+p} \sqrt{1-\bold R^{n+p}}\frac{(n+p) \left(n N \sinh ^2\alpha+n+p+1\right)}{2-\bold R^{n+p}(n+p+2)}~~,
\eeq 
where we have defined $V_{(p)}=\Omega_{(p)}R^{p}$. We note that in $D=5$ and for a ring geometry ($p=1$), that is, $n=1$, and in the uncharged limit ($\alpha=0$), this describes the black ring surrounding the spherical black hole in the black saturn solution of \cite{Elvang:2007rd}. In App.~\ref{app:saturn} we show explicitly, by taking the blackfold limit of \cite{Elvang:2007rd}, that this is indeed the case.\footnote{In the charged case for Kaluza-Klein coupling ($N=1$) this potentially describes a particular limit of the charged black saturn solution obtained in \cite{Grunau:2014vwa}. We leave this check for future work.} We note that the thermodynamic properties of this solution and considering the tension per unit length ${\boldsymbol{\mathcal{T}}}=\hat{\boldsymbol{\mathcal{T}}}/\mu$, it is straightforward to check that the first law \eqref{eq:1stlaw2} is satisfied and, furthermore, both ${\boldsymbol{\mathcal{T}}}$ and $\hat{\boldsymbol{\mathcal{T}}}$ vanish when $\mu\to0$.

\subsubsection*{Spacetime pressure and black hole volume in black hole backgrounds}
As we have shown in the introduction, these solutions satisfy the first law \eqref{eq:1stlaw} for an infinite set of quantities. We may introduce a notion of black hole volume by choosing $k=-2$, which leads to the black hole volume
\beq \label{eq:volschz}
V_i=\frac{\Omega_{(n+1)}V_{(p)}}{32\pi G}\frac{r_0^{n}}{\lambda}\mu^2\bold{R}^{n+p} \sqrt{1-{\bold{R}}^{n+p}}\frac{(n+p) \left(n N \sinh ^2\alpha+n+p+1\right)}{2-{\bold{R}}^{n+p}(n+p+2)}~~.
\eeq
From here we see that once we remove the black hole $\mu\to0$, the volume \eqref{eq:volschz} vanishes. This is not the expected result for the volume in this limit. More precisely, the volume for the black ring ($p=1$) has been computed in \cite{Altamirano:2014tva} by taking the flat spacetime limit of the perturbative construction of an uncharged AdS black ring in the ultraspinning regime \cite{Caldarelli:2008pz}. The volume for this case scales like $V_i\propto r_0^{n}R^2$ in the flat spacetime limit. Since this solution corresponds to the case $\mu\to0$ and $\alpha=0,p=1$ in \eqref{eq:volschz}, we see that the volume introduced in \eqref{eq:volschz} does not scale in the expected way. We could introduce by hand a new length scale $\tilde L$ such that $(\tilde L/R)^2=(\mu/R)^{D-3}$ which would lead to the right scaling in the flat spacetime limit. However, this would not satisfy the first law \eqref{eq:1stlaw} with a pressure of the form $P_e=\lambda \tilde L^{-2}$, even if we would try to define a new mass $\tilde M$ as in Sec.~\ref{sec:bhc}. Again, therefore, we conclude that the notion of gravitational tension is more natural than the notion of black hole volume.

\subsubsection*{Blackfold mass in asymptotically flat black hole backgrounds}
We note that the definition of gravitational tension is related to the generalised first law of thermodynamics \eqref{eq:1stlaw2}, and hence to a specific definition of mass/energy, namely, the definition of mass that enforces \eqref{eq:1stlaw2} for a given black hole horizon in the presence of surrounding gravitational fields. When using the free energy \eqref{eq:freeodd} and the thermodynamic relations \eqref{eq:thermo},\eqref{eq:freerelation} we obtain the mass of this blackfold construction, which is given by
\beq \label{eq:masssch}
M=\frac{\Omega_{(n+1)}V_{(p)}}{8\pi G}r_0^{n} (1-\bold{R}^{n+p})^{\frac{3}{2}} \frac{ \left(n N \sinh ^2\alpha+n+p+1\right)}{2-\bold{R}^{n+p}(n+p+2)}~~.
\eeq
However, as we show in App.~\ref{app:saturn} this mass, with the appropriate values of the several constants involved, does not correspond to the Komar mass $M^{\text{BR}}$ measured near the black ring horizon of the black saturn solution. In fact we find that 
\beq \label{eq:massK}
M^{\text{BR}}=M-\frac{\hat{\boldsymbol{\mathcal{T}}}}{(D-3)}~~.
\eeq
This is expected since the authors of \cite{Elvang:2007rd} have shown that the first law \eqref{eq:1stlaw} holds for the black saturn solution where $M$ is the total ADM mass, which is the sum of the Komar masses of the centre black hole and of the black ring. However, if we require the existence of a first law that holds for each individual horizon then we must introduce a new mass measured in connection with each separate horizon. For the black ring horizon, this is precisely \eqref{eq:masssch}, that is, the mass that is directly obtained from the blackfold approach, which already takes into account the contribution from the gravitational binding energy due to the presence of the centre black hole. Note that this is rather different than the KK case of Sec.~\ref{sec:bhc}, since there both the original mass $M$ and the shifted mass $\tilde M$ satisfied the first law of thermodynamics. In this case, only the blackfold mass \eqref{eq:masssch}, obtained from general principles, satisfies the first law.

\subsubsection*{Free energy for the black ring in the black saturn solution}
From the above discussion about the mass we are lead to an intriguing consequence for the thermodynamics of disconnected horizons, namely, the free energy for a given horizon, obtained via local computations of the conserved quantities, does not behave in a thermodynamically correct way. 

In the case of the black saturn solution in $D=5$, if we denote the black ring mass and angular momentum, which can be obtained via Komar integrations near the horizon \cite{Elvang:2007rd}, by $M^{BR}$ and $J^{BR}$ , and furthermore, denoting the black ring horizon temperature, angular velocity and entropy by $T^{BR}$, $\Omega^{BR}$ and $S^{BR}$, then the free energy $\mathcal F^{BR}=M^{BR}-T^{BR}S^{BR}-\Omega^{BR}J^{BR}$ can be seen as the free energy of a black ring in a non-trivial background for which the background length scale is that associated with the mass $M^{BH}$ of the black hole in the center. However, we find that the following expected thermodynamic relations do not hold, i.e.,
\beq \label{eq:thermobr1}
S^{BR}|_{M^{BH},\Omega^{BR}}\ne-\frac{\partial \mathcal F^{BR}}{\partial T^{BR}}~~,~~J^{BR}|_{M^{BH},T^{BR}}\ne-\frac{\partial \mathcal F^{BR}}{\partial \Omega^{BR}}~~.
\eeq
Indeed this could have been anticipated, due to the fact that in the ultraspinning limit, the black ring mass, obtained from the blackfold approach, does not coincide with the Komar integration on the black ring horizon. Therefore, in order to define a proper free energy for a given horizon in a black hole solution with multiple disconnected horizons, we must introduce another notion of mass for that specific horizon.

We now give an explicit construction of this mass for the black ring in the black saturn solution for which the angular momentum of the center black hole vanishes. All thermodynamic quantities can be parametrised by 3 parameters $L,\beta,k_2$ (see App.~\ref{app:saturn}), where $\beta$ controls the mass of the center black hole. If $\beta=0$ we obtain the pure black ring solution. The relation between these parameters, in the ultraspinning limit, and those used to parametrize the solution \eqref{eq:solscz}, is given in \eqref{eq:cond}-\eqref{eq:cond22}. We now proceed by introducing a new mass $M$ by adding to the Komar mass $M^{BR}$ a term proportional to a function $f(\beta,k_2)$ such that
\beq \label{eq:mnew}
M=\frac{3\pi}{4G}L^2\left(k_2+f(\beta,k_2)\right)~~,
\eeq
and we want to demand that the resulting free energy $\mathcal F=M-T^{BR}S^{BR}-\Omega^{BR} J^{BR}$ satisfies the thermodynamic relations \eqref{eq:thermobr1}. Due to the cumbersome expressions inherent to the black saturn solution, we will show how this is done to first order in the ultraspinning approximation, i.e., in an expansion around $k_2=0$. We first decompose $f(\beta,k_2)$ as
\beq
f(\beta,k_2)=k_2f(\beta)_{(0)}+k_2^2f(\beta)_{(1)}+\mathcal{O}(k_2^{3})~~.
\eeq
To leading order in the ultraspinning limit, there is already a correction, as we have seen above, which one can easily corroborate from the analytic solution and hence obtaining $f(\beta)_{(0)}=\beta/(\beta-2)$. With this value of $f(\beta)_{(0)}$, the mass \eqref{eq:mnew} to order $\mathcal{O}(k_2^2)$ yields \eqref{eq:masssch} in the uncharged case ($\alpha=0$), as shown in App.~\ref{app:saturn}. Proceeding to next order we find,
\beq
f(\beta)_{(1)}=\frac{\beta\left(\beta(17+\beta(2\beta-5))-12\right)}{6(\beta-2)^3(\beta-1)}~~.
\eeq
Now, using the Smarr relation \eqref{eq:smarr2} we find the total tension
\beq
\hat{\boldsymbol{\mathcal{T}}}=-\frac{3\pi L^2}{2G}k_2\beta\left(\frac{1}{(2-\beta)}+\frac{\left(\beta(17+\beta(2\beta-5))-12\right)}{6(\beta-2)^3(1-\beta)}k_2+\mathcal{O}(k_2^2)\right)~~,
\eeq
which when using \eqref{eq:cond}-\eqref{eq:cond22}, to leading order in $k_2$, coincides with \eqref{eq:tensch}. This procedure can be iteratively continued to arbitrary orders in $k_2$. It would be interesting to obtain an exact expression for the shifted mass to all orders, and consequently obtain the exact gravitational tension for the black ring in the black saturn solution. Furthermore, a similar analysis can be carried out for the free energy of the centre black hole horizon, which can be thought of as a black hole placed in the gravitational field of a black ring. We leave this interesting analysis for future work.

\subsection{Schwarzschild-(A)dS black hole background}
We now briefly consider a similar construction in (A)dS spacetimes by placing a ring surrounding a Schwarzschild-(A)dS black hole. The metric of the Schwarzschild-(A)dS black hole takes the same form as in \eqref{eq:dsads} but with the blackening factor
\beq\label{eq:fschAdS}
f(r) &=& 1+\frac{r^2}{L^2}-\left(\frac{\mu}{r}\right)^{n+p}~~,
\eeq
where $\mu$ is the horizon location of the Schwarzschild black hole and $L$ is the size of (A)dS. In this background we place a neutral odd-sphere geometry at a fixed $r=R$ and hence the induced metric and Killing vector field are the same as in Sec.~\ref{sec:oddads} as well as the free energy \eqref{eq:freeodd}, but with the function \eqref{eq:fschAdS}. The equilibrium angular velocity is given by \eqref{eq:equigen} and using \eqref{eq:fschAdS} we can rewrite the equilibrium condition as 
\be
\!\!\!\!\Omega^2 R^2=\frac{\left(1+\textbf{R}^2\!-\!\tilde{\textbf{R}}^{n+p}\right) \left((n+1) \left(2\textbf{R}^2\!+\!(n+p)\tilde{\textbf{R}}^{n+p}\right)+2 p
   \left(1\!+\!\textbf{R}^2\!-\!\tilde{\textbf{R}}^{n+p}\right)\right)}{ \left(\left(2\textbf{R}^2+(n+p)
   \tilde{\textbf{R}}^{n+p}\right)+2 (n+p) \left(1+\textbf{R}^2-\tilde{\textbf{R}}^{n+p}\right)+2 p
   \right)}~~,
\ee
where we have defined the dimensionless quantities $\textbf{R}=R/L$ and $\tilde{\textbf{R}}=(\mu/R)$. We now proceed and analyse the total integrated tension $\hat{\boldsymbol{\mathcal T}}$ from \eqref{eq:freeoddpp}, which is given as sum of tensions, one for the black hole and one for the the (A)dS radius. The results are the following
\beq
\hat{\boldsymbol{\mathcal T}}_{\mu} &=& \frac{V_{(p)}\Omega_{(n+1)}}{16\pi G}r_0^n\tilde{\textbf{R}}^{n+p}\frac{(n+p+1)(n+p) \sqrt{1+\textbf{R}^2-\tilde{\textbf{R}}^{n+p}}}{(n+p+2)\tilde{\textbf{R}}^{n+p}-2}~~,\\
\hat{\boldsymbol{\mathcal T}}_L &=&\frac{V_{(p)}\Omega_{(n+1)}}{8\pi G}\textbf{R}^2\frac{ (n+p+1) \sqrt{1+\textbf{R}^2-\tilde{\textbf{R}}^{n+p}}}{(n+p+2) \tilde{\textbf{R}}^{n+p}-2}r_0^n~~.
\eeq
Since each scale is independent, we can view this spacetime as having two distinct tensions: the one associated with AdS spacetime and the one associated with Schwarzschild spacetime. By defining the corresponding tensions per unit length, it is straightforward to see that the first law \eqref{eq:1stlaw2} holds. The remaining thermodynamic properties can also be easily obtained as in the previous examples.


\section{Non-universality of black hole volume}\label{sec:non}
In this section we analyse the consequences of the existence of the perturbative solutions of Sec.~\ref{sec:bf} to the notion of black hole volume. We have already argued that this notion was unnatural and did not lead to he expected scaling for both the KK black holes of Sec.~\ref{sec:bhc} and the higher-dimensional black saturn solutions of Sec.~\ref{sec:sch}. However, since a sceptic reader might consider those examples too exotic or far removed from the case of black holes in AdS, we focus in this section on two simpler examples. In particular, we look at the family of (charged) black odd-spheres of Secs.~\ref{sec:oddads}, \ref{sec:planeodd} and \ref{sec:lifs} in AdS, plane wave and Lifshitz spacetimes respectively, as well as the family of charged rotating black holes in Secs.~\ref{sec:discads} and \ref{sec:discpp} in AdS and plane wave spacetimes respectively.

Our methodology consists in analysing the flat spacetime limit of each of these families of solutions. Since in this limit all the different cases within each familiy reduce to the same flat spacetime black hole solution, then, if the notion of black hole volume is to be meaningful in this limit, it must be universal. If we denote the black hole volume for a given family of solutions in the flat spacetime limit by $\tilde V_i$, then we must require that
\beq\label{eq:limit}
V_i|_{L_a\to\infty}\to \tilde V_i~~.
\eeq
However, as we will see, if we require this to be the case for the family of black odd-spheres (which contains black rings as a particular case) then we are forced to accept that black odd-spheres in flat spacetime are characterized by an infinite set of volumes. Furthermore, we will see that it is not possible to demand \eqref{eq:limit} for the family of charged black discs (charged rotating black holes).

\subsection{Black odd-spheres}
We first consider the case of the black odd-spheres in AdS, which we constructed in Sec.~\ref{sec:oddads}. Since we are in AdS, we take \eqref{eq:pads} as a working definition of pressure in AdS. Therefore we find the black hole volume
\beq \label{eq:oddvolads}
V_i=\frac{ \Omega_{(n+1)}V_{(p)}}{2 (n+p+1) (n+p+2)}r_0^nR^2 \sqrt{1+\bold R^2} \left(nN\sinh^2\alpha+n+p+1\right)~~,
\eeq
for this particular class of AdS solutions. In the uncharged case ($\alpha=0$) and for the ring ($p=1$) this volume had been obtained in \cite{Altamirano:2014tva}. From here we obtain the non-zero flat spacetime limit of the volume \eqref{eq:oddvolads} by taking $\bold R\to0$. For simplicity, focusing on the uncharged case $\alpha=0$, we obtain
\beq \label{eq:flatodd}
\tilde V_i=\frac{ \Omega_{(n+1)}V_{(p)}}{2 (n+p+2)}r_0^nR^2 ~~.
\eeq
We now consider, by the same token, obtaining the black hole volume for the class of Lifshitz solutions of Sec.~\ref{sec:lifs}. We note that in Lifshitz spacetimes, the authors of Ref.~\cite{Brenna:2015pqa} have recently proposed the following definition of the spacetime pressure
\beq\label{eq:plif}
P_e=\frac{(D+z-2)(D+z-3)}{16\pi L^2}~~,
\eeq
which reduces to \eqref{eq:pads} when $z=1$. Using \eqref{eq:plif}, the black hole volume for this class of solutions is given by
\beq
V_i=\frac{\Omega_{(n+1)}}{2}V_{(p)}r_0^{n} \frac{\sqrt{\alpha +\bold R^2} \left(z \left(\beta +\bold R^2\right)-\beta \right) \bold R^{z-1}}{(n+p+z+1) \left|\bold R^2
   (z-1)+\beta  (z-2)\right|}~~.
\ee
This indeed reduces to the volume \eqref{eq:oddvolads} when $z=1$ and hence has the same flat spacetime limit \eqref{eq:flatodd}. This corroborates the choice of spacetime pressure \eqref{eq:plif} by the authors \cite{Brenna:2015pqa}.

However, let us consider the case of the odd-spheres in plane wave spacetimes constructed in Sec.~\ref{sec:planeodd}. These solutions reduce to the same flat spacetime odd-spheres as the previous two cases when $A_{\hat a}\to0$. Since in plane wave spacetimes we have $m$ length scales and each length scale is independent, then, associated with each tension computed in \eqref{eq:tensionpp}, we have a specific black hole volume $V^{\hat a}_i$. Taking the pressure on each direction to be given by
\beq \label{eq:ppp}
P_e^{\hat a}=\frac{(D-1)(D-2)}{16\pi}A_{\hat{a}}~~,
\eeq
we obtain a set of black hole volumes
\beq
V^{\hat a}_i=\frac{ \Omega_{(n+1)}V_{(p)}}{2 (n+p+2)}r_0^nR_{\hat a}^2 \sqrt{1+\bold R^2}~~,
\eeq
each associated with one of the $p_{\hat a}$-spatial parts of the worldvolume. The existence of this set of volumes is a direct consequence of the anisotropy of the spacetime. The definition of pressures \eqref{eq:ppp} is indeed the correct one for these configurations since the flat spacetime limit of the volume ($A_{\hat a}\to0$) coincides with \eqref{eq:flatodd}. However, since we have an arbitrary number of volumes $V^{\hat a}_i$ associated with each non-trivial plane wave direction, then we must conclude that the corresponding flat spacetime black holes would be characterised by an arbitrary number of black hole volumes. For this reason, we argue that it is more natural to work with the gravitational tension associated with each spacetime direction, all of which vanish when taking the flat spacetime limit. 

\subsection{Charged rotating black holes}
We first consider the analogue of the Kerr-Newman solution in higher-dimensions in AdS constructed in \ref{sec:discads}. Using the results of that section we obtain black hole volume
\be\label{eq:volads}
\begin{split}
\left(V_i\right)_{\text{AdS}}&=  \frac{2\Omega_{(n+1)}}{(D-1)(D-2)}\tilde r_0^n\frac{\,
   _2F_1\left(1,\frac{1}{2} (N-1) n;\frac{Nn}{2}+2;1-\frac{\Phi
   ^2_{\text{H}}}{N}\right)}{(2+N n) \xi^2\Omega^4}~~,
 \end{split}
\ee
where we have used the definition of pressure \eqref{eq:pads}. This volume in the limit $L\to\infty$ and for $\Phi_{\text{H}}=0$ yields the black hole volume corresponding to the Myers-Perry black hole in the ultraspinning regime in $D\ge6$, which has been analysed in \cite{Cvetic:2010jb}. In the uncharged case and in (A)dS, this volume has also been computed in \cite{Cvetic:2010jb, Dolan:2013ft, Altamirano:2014tva} for the entire family of higher-dimensional Kerr-(A)dS black holes. The exact form obtained here for the ultraspinning regime was only analysed in \cite{Altamirano:2014tva} and for the special case where $\xi\to0$ for which the black hole saturates the BPS bound \cite{Chrusciel:2006zs} in AdS.

We now consider the case of the charged black holes in plane wave spacetimes constructed in Sec.~\ref{sec:discpp} which have the same flat spacetime limit as the case above. In this case we take the pressure to be of the form $P_e=\lambda A_1 G^{-1}$. Using \eqref{eq:tensiondiscpp}, we express the black hole volume in terms of $\lambda$ and the tension $\hat{\boldsymbol{\mathcal{T}}}$ such that
\beq \label{eq:volpp}
\left(V_i\right)_{\text{pp}}=-\frac{G\hat{\boldsymbol{\mathcal{T}}}}{2\lambda A_1}~~.
\eeq
In order to obtain the correct factor $\lambda$ we compare the volume \eqref{eq:volads} in the flat spacetime limit $L\to\infty$ and the volume \eqref{eq:volpp} in the same flat spacetime limit $A_1\to0$. We find the ratio
\beq
\frac{(\tilde V_i)_{\text{AdS}}}{(\tilde V_i)_{\text{pp}}}=\frac{16\pi \lambda}{(n+3)(n+5-\Phi^2_{\text{H}})}~~.
\eeq
If we demand the black hole volume to be universal in the flat spacetime limit we must require the above ratio to be equal to unity. However, we would have to require $\lambda$ to have a dependence on the chemical potential $\Phi_{\text{H}}$ which is not possibe since $\lambda$ is a constant by definition. If we were to allow such dependence then the first law \eqref{eq:1stlaw} would bot be satisfied. Therefore, we conclude, thermodynamic black hole volume in the flat spacetime limit is not universal. This indeed suggests that the definitions of black hole volume introduced in the literature \cite{Parikh:2005qs,Ballik:2010rx,Ballik:2013uia} are not fundamentally related to the thermodynamic black hole volume.


\section{Outlook \label{sec:out} } 
In this paper we have shown that there is an infinite set of conjugate thermodynamic variables $(B_k,\mathbb{L}_k)$ that can be introduced in order to describe the modifications in the first law \eqref{eq:1stlaw} due to variations of external gravitational fields. We have argued that the most natural quantity that describes these variations is the gravitational tension (or gravitational binding energy) that describes the extra energy associated to a black hole due to the presence of surrounding gravitational fields. We have furthermore argued that the popular choice of black hole volume and spacetime pressure used to describe such variations in AdS spacetimes is not the most natural one and leads to non-universal results in the flat spacetime limit. 

In order to reach these conclusions we have proposed in Sec.~\ref{sec:intro} that modifications to \eqref{eq:1stlaw} should satisfy four different properties. We could, in principle, not demand property \textbf{(4)}, namely, the existence of a universal result when the external gravitational field is removed. Imposing it, selects the introduction of gravitational tension instead of black hole volume to describe the modifications of \eqref{eq:1stlaw}, namely, via \eqref{eq:1stlaw2}. Not imposing it, would in principle render any of the choices of $k$ in \eqref{eq:1stads} as good as any other. However, we must also recall property \textbf{(1)}, which requires the existence of a geometrical interpretation. Since we have shown that the notion of black hole volume can be defined in spacetimes which are solutions of the vacuum Einstein equations, then the geometrical interpretation in terms of Killing potentials \cite{Kastor:2009wy} does not hold. On the other hand, there is a well-defined prescription for evaluating the gravitational tension, following \cite{Harmark:2004ch}, that works for arbitrary black hole spacetimes, at least when there are periodic or non-compact horizon directions or whenever the black hole admits a blackfold regime. In order to complete this picture, we would need generalise the prescription of \cite{Harmark:2004ch} to black holes which do not admit a blackfold regime. This interesting task we leave for future work.

Furthermore, when considering complete UV theories of gravity, we must in fact add an extra property, namely, \textbf{(5)} the existence of a microscopic description. As it is well known, the entropy of black holes has played a central role in developing and testing theories of quantum gravity. In particular, it has led to the celebrated holographic principle \cite{'tHooft:1993gx,Susskind:1994vu} as embodied in the AdS/CFT correspondence \cite{Aharony:1999ti}. The existence of a macroscopic entropy poses the challenge of a microscopic explanation,
and one of the successes of string theory has been to provide this for classes of supersymmetric black holes \cite{Strominger:1996sh}. Similarly, one may expect that such microscopic description should also exist for the quantities describing variations in the gravitational field. In the particular case of AdS, recent work \cite{Karch:2015rpa} has given a possible CFT interpretation of black hole volume. However, we think that a similar interpretation could be given for gravitational tension and this research direction would be very interesting to pursue. 

We conclude with some interesting observations and future research directions.

\subsubsection*{A van der Walls interpretation}
We would also like to comment on the van der Waals interpretation \eqref{eq:pads} in terms of the quantities $(\mathbb{B}_k,\mathbb{L}_k)$ for general $k$. In fact, this interpretation relies only on a rewriting of the horizon radius $r_+$ in terms of the the quantities $(\mathbb{B}_k,\mathbb{L}_k)$, including $k=-2$, for which the specific volume $v$ in \eqref{eq:pads} is given by $v=2r_+$. In fact, using \eqref{eq:master} for the Schwarzschild-AdS black hole we find
\beq \label{eq:bkk}
\mathbb{B}_k=-\frac{r_+^3}{k\lambda L^{k+2}}~~,
\eeq
which is a result that also holds for the Reisnner-Nordstr\"{o}m-AdS black hole (see e.g. \cite{Kubiznak:2012wp} for the explicit thermodynamic quantities). Using \eqref{eq:bkk} we can introduce what is usually referred to as specific volume $v$ \cite{Kubiznak:2012wp}
\beq \label{eq:spec}
v=2(-k\lambda^{-\frac{2}{k}} \mathbb{L}_k^{\frac{2+k}{k}} \mathbb{B}_k)^{\frac{1}{3}}=2 r_+~~,
\eeq
where the $k$-dependence in the second expression has been chosen in order to make $v$ $L$-independent (with $k\ne0$). The resulting equation of state $\mathbb{L}_k=\mathbb{L}_k(T,Q,v)$ for the Reisnner-Nordstr\"{o}m-AdS black hole is then
\beq
\mathbb{L}_k^{\frac{2}{k}}=\frac{3\lambda ^{\frac{2}{k}}}{4}\frac{v^{2}}{2\pi vT+\frac{Q^2}{v^2}-1}~~,
\eeq
where $Q$ is its electric charge. If one chooses $k=-2$ then this equation takes a form that resembles a van der Waals-type equation and in general it can be seen as a van der Waals-type equation to some power dependent on $k$. We may obtain the critical points for any $k$ by requiring
\beq
\left(\frac{\partial \mathbb{L}_k}{\partial v}\right)_{T,Q}=0~~,~~\left(\frac{\partial^2 \mathbb{L}_k}{\partial v^2}\right)_{T,Q}=0~~.
\eeq
This leads to $(\mathbb{L}_k)_c=\lambda 6^{k}Q^{k}$ as well as to the universal result for any value of $k\ne0$, namely,
\beq
T_c=\frac{\sqrt{6}}{18\pi Q}~~,~~v_c=2\sqrt{6}Q~~,
\eeq
which had been previously obtained in \cite{Kubiznak:2012wp} for $k=-2$. Therefore, we are not surprised that the work of \cite{Caceres:2015vsa} observed a transition in the entanglement entropy for several charge configurations, in the context of STU black holes, at a critical temperature which was computed using $k=-2$ since the critical temperature is independent of the value of $k$ that one selects. 

\subsubsection*{First law for black holes with multiple disconnected horizons}
In Sec.~\ref{sec:bh} we have studied black saturn configurations in both flat and AdS spacetimes. In flat spacetime, the first law of black hole mechanics was derived in \cite{Elvang:2007hg} for multiple disconnected horizons and relates variations of the total mass $M^T$ of the combined system to variations of the entropy and angular momentum measured near each individual horizon, in particular,
\beq \label{eq:total1st}
dM^T=\sum_{\hat i}\left(T^{\hat i}dS^{\hat i}+\Omega^{\hat i} dJ^{\hat i}\right)~~,
\eeq
where the index $\hat i$ runs over each disconnected horizon. In vacuum flat spacetime, the total mass $M^T$ is given by the sum of the individual Komar masses, in this case $M^{\text{BH}}$ and $M^{\text{BR}}$. Using this into \eqref{eq:total1st} one can trivially define first laws for each individual horizon by moving the contribution of one of the Komar masses to the r.h.s., however, as shown in Sec.~\ref{sec:bh}, this notion of mass is not consistent with the notion of free energy $\mathcal{F}^{\hat i}=M^{\hat i}-T^{\hat i}S^{\hat i}-\Omega^{\hat i}J^{\hat i}$ defined for each individual horizon. To reconcile this, we have argued that the appropriate notion of mass defined for each individual horizon is the one that takes into account the gravitational binding energy due to the presence of the other horizon.\footnote{We have given evidence in Sec.~\ref{sec:bh} that this argument also holds in AdS backgrounds.} This mass, in the ultraspinning limit, is the mass computed from first principles using the blackfold approach and we have shown that it is possible to obtain it from the exact black saturn solution to higher orders in the ultraspinning limit. 

It would be worthwhile obtaining this mass, that takes into account the binding energy, exactly (i.e. to all orders in the blackfold expansion) for the black saturn solution. We think that this perspective, namely, of defining individual first laws to each disconnected horizon, is a perspective that can contribute to the understanding of the thermodynamics of disconnected horizons and be applicable to cases where these are inherently present, such as in the case of asymptotically deSitter black holes. Furthermore, these cases are examples of cases satisfying the first law \eqref{eq:1stads} but where variations associated with the background length scales do not change the theory. In contrast to variations of the cosmological constant, variations due to the black hole mass do not change the theory as they are controlled by parameters in the solution and not in the Lagrangian. Therefore, toy models of black hole binary systems provide an excellent testing ground for new thermodynamic behaviour. 

\subsubsection*{New black hole solutions}

In this paper, we have constructed novel perturbative black hole solutions in AdS, plane wave and Lifshitz spacetimes. In particular we constructed the analogue of the higher-dimensional Kerr-Newman solution in AdS and we gave the first example of a class of black hole solutions in Lifshitz spacetimes that is stationary and furthermore has non-trivial horizon topology. These solutions were found using the blackfold approach and their consequences for the universality of black hole volume in the flat spacetime limit were studied. This exercise is yet another illustration of how the blackfold approach can be used in a very simple way to study interesting properties of higher-dimensional black holes. 

In connection with the novel Lifshitz solutions that we have obtained, we note that since Lifshitz spacetimes play a role in holography for field theory
systems with anisotropic scaling between time and space \cite{Kachru:2008yh, Taylor:2008tg}, it would be interesting to generalise the perturbative Lifshitz solutions of this paper in order to include charge. In particular, an interesting family would be the analogue of the Kerr-Newman solutions of Secs.~\ref{sec:discads}, \ref{sec:discpp} in Lifshitz spacetimes. These would be the rotating versions of the black holes constructed in \cite{Tarrio:2011de}. Since for these solutions the Lifshitz vacuum is supported by non-trivial matter fields, this would require the use of the generalisation of the blackfold equations of motion to such backgrounds \cite{Armas:2016mes}. 

\subsubsection*{Other backgrounds and Taub-Nut-AdS/Taub-Bolt-AdS black holes}

We also briefly mention a number of more general settings in which it would be interesting to examine the gravitational tension perspective proposed in this paper.
These include Taub-NUT spacetimes\footnote{Black hole volume and spacetime pressure have been considered for such spacetimes in e.g. \cite{Johnson:2014yja}.}, KK bubbles \cite{Witten:1981gj} and the more general sequences of bubbles and holes (see e.g. \cite{Elvang:2004iz}), other limits of the black saturn configuration, black holes in flux bagrounds, and high-derivative gravity theories.\footnote{See e.g. \cite{Cai:2013qga} for RN-AdS black holes in Gauss-Bonnet gravity.} As a proof of concept and to finalise this discussion, we quickly consider the case of $D=4$ uncharged black holes in Taub-Nut-AdS (TN) and Taub-Bolt-AdS (TB) spacetimes studied in \cite{Johnson:2014pwa}. The free energy in these cases can be written as \cite{Johnson:2014pwa}
\beq
\mathcal{F}=\frac{4\pi N}{L^2}\left(L^2m^2+3N^2r_+-r_+^3\right)~~,
\eeq
where $N$ is the Nut parameter, related to the temperature  $T$ via $T=1/(8\pi N)$, $m$ is the mass parameter and $r_+$ the horizon radius. For each of the two cases, the mass parameter $m$ and the horizon radius $r_+$ are determined in terms of $N$ and $L$ \cite{Johnson:2014pwa} and hence the free energy can be expressed as a function of $T$ and $L$ . We consider applying formula \eqref{eq:master} with $k=1$, i.e. $\mathbb{L}_{1}=L$. We find the respective tensions
\beq \label{eq:taub}
\boldsymbol{\mathcal{T}}_{\text{TN}}=\frac{2N^3}{GL^3}~~,~~\boldsymbol{\mathcal{T}}_{\text{TB}}=-\frac{15N^3}{2G L^3}+\mathcal{O}\left(\frac{N^5}{L^5}\right)~~,
\eeq
where, for simplicity, we have given the result for Taub-Bolt-AdS for small Nut parameter $N$ (or equivalently large AdS radius $L$). These black holes satisfy, as all previous examples, the first law \eqref{eq:1stads} with their respective tensions. Furthermore, they exhibit the correct universal limit of the tension in flat spacetime, namely, if either $N\to0$ or $L\to\infty$ the tension vanishes.\footnote{In the case of Taub-Bolt-AdS there are two branches of black hole solutions \cite{Johnson:2014pwa}. The $(-)$ branch, which was the one used in \eqref{eq:taub}, connects to flat spacetime when $L\to\infty$. The $(+)$ branch is analogous to large black holes in AdS and hence does not have a flat space counterpart. For this reason it cannot be used to argue for a universal limit of the tension in the flat space limit. However, the tension can also be introduced in such case (for small $N$ it has the behaviour $\boldsymbol{\mathcal{T}}_{\text{TB}}\sim L^3/N^3$) and the first law \eqref{eq:1stads} is satisfied.} This provides good evidence that all black hole spacetimes can be treated within the framework developed here.

 \section*{Acknowledgments}
We would like to acknowledge important discussions with Roberto Emparan, Troels Harmark, Jelle Hartong, Cynthia Keeler and Vasilis Niarchos. We also thank the referee for useful comments on the paper which have been incorporated in the final version. JA acknowledges the current support of the ERC Starting Grant 335146 HoloBHC. JA is grateful to the COST network for a short term mission grant.  The work of NO and MS is supported in part by the Danish National Research Foundation project ``New horizons in particle and condensed matter physics from black holes".

\appendix

\section{Thermodynamic properties of blackfold configurations} \label{app:thermo}
Here we collect and present most of the thermodynamic properties of the configurations found in the core of this paper. 


\subsubsection*{Electrically charged black odd-spheres in (A)dS}
The thermodynamic properties of the configurations found in Sec.~\ref{sec:oddads} can be easily obtained from \eqref{eq:freeodd} using formulae \eqref{eq:thermo},\eqref{eq:freerelation} and written in terms of $r_0, R, \alpha$. These read
\be\begin{split}
M=&\frac{\Omega_{(n+1)}V_{(p)}}{16\pi G}r_0^n\left(1+\bold R^2\right)^{3/2} \left(n N  \sinh^2\alpha+n+p+1\right)~~,\\
J=&\frac{\Omega_{(n+1)}V_{(p)}}{16\pi G}r_0^{n}R\frac{ \sqrt{\left(\bold R^2 (n N  \sinh^2\alpha+n+p+1)+p\right) }}{n }\\
&\times \frac{\sqrt{\left((n N\sinh^2\alpha+n+p)(1+\textbf{R}^2)+\textbf{R}^2\right)}}{n}~~,\\
S=&\frac{\Omega_{(n+1)}V_{(p)}}{4G}r_0^{n+1}  \cosh ^{ N}\alpha\sqrt{\frac{ \bold R^2\left (n N \sinh ^2\alpha+n+p+1\right)+n N \sinh ^2\alpha+n+p}{n \left(1+N \sinh ^2\alpha\right)}}~~,\\
Q=&\frac{\Omega_{(n+1)}V_{(p)}n}{16\pi G}\sqrt{N} g(\alpha) r_0^{n}\sqrt{\frac{\bold R^2\left (n N \sinh ^2\alpha+n+p+1\right)+n N \sinh ^2\alpha+n+p}{n \left(1+N \sinh ^2\alpha\right)}}~~,
\end{split}\ee
where $g(\alpha)=\sinh \alpha\cosh \alpha$ while the temperature $T$ and chemical potential $\Phi_{\text{H}}$ are given in terms of $r_0, R, \alpha$ via the expressions
\be
\begin{split}
T&= \frac{n}{4 \pi  r_0}\sqrt{\frac{n \left(\bold R^2+1\right) \left(N   \sinh ^2\alpha+1\right)}{ \bold R^2\left(n N   \sinh ^2\alpha +n+p+1\right)+n N \sinh ^2\alpha+n+p}}~~,\\
\Phi_{\text{H}}&=\tanh\alpha\sqrt{\frac{n \left(\bold R^2+1\right) \left(N   \sinh ^2\alpha+1\right)}{ \bold R^2\left(n N   \sinh ^2\alpha +n+p+1\right)+n N \sinh ^2\alpha+n+p}}~~.
\end{split}
\ee


\subsubsection*{Black discs in (A)dS: analogue of Kerr-Newmann black holes}
For these black holes found in Sec.~\ref{sec:discads}, the thermodynamic properties can be extracted from \eqref{eq:freedisc} and read
\be\begin{split}
J=&\frac{\Omega_{(n+1)}}{4G}\tilde r_0^{n}\frac{ \, _2F_1\left(1,\frac{1}{2} n (N-1);\frac{n
   N}{2}+2;1-\frac{\Phi ^2_{\text{H}}}{N}\right)}{\Omega^3 \xi^2(2+n N)}~,\\
S=&\frac{n\Omega_{(n+1)}}{8TG}\tilde r_0^n\frac{  \,
   _2F_1\left(1,\frac{1}{2} n (N-1);\frac{n N}{2}+2;1-\frac{\Phi
   ^2_{\text{H}}}{N}\right)}{\Omega^2\xi (2+n N)}~,\\
Q=&\frac{\Omega_{(n+1)}}{32G}\frac{n \Phi_{\text{H}}  \tilde r_0^n \Gamma
   \left(\frac{n N}{2}\right)}{\Omega^2\xi\left(1-\frac{\Phi_{\text{H}}^2}{N}\right)} \Big(n N \, _2\tilde{F}_1\left(1,\frac{1}{2} n
   (N-1);\frac{n N}{2}+2;1-\frac{\Phi_{\text{H}} ^2}{N}\right)\\
   &+2 \,
   _2\tilde{F}_1\left(2,\frac{1}{2} n (N-1);\frac{n N}{2}+2;1-\frac{\Phi_{\text{H}}
   ^2}{N}\right)\Big)~~.
\end{split}\ee
The mass of the black hole can be readily obtained using \eqref{eq:freerelation} together with the free energy \eqref{eq:freedisc} and the above quantities.


\subsubsection*{Electrically charged black discs in plane wave background}
Here we collect the thermodynamic properties of the configurations of Sec.~\ref{sec:discpp}. These are given by
\be\begin{split}
M=&\frac{\Omega_{(n+1)}}{8 G}\tilde r_0^n\frac{(n+2) \left(\Omega ^2-A_1\right)-\left(\Phi ^2_{\text{H}}-1\right) \left(2
   A_1+\Omega ^2\right) \,
   _2F_1\left(-\frac{1}{2},1;\frac{n+4}{2};\frac{A_1 (\Phi_{\text{H}}^2 -1) }{\Omega ^2-A_1}\right)}{(n+2)(1-\Phi ^2_{\text{H}}) \left(A_1-\Omega ^2\right)^2}~,\\
J=&\frac{\Omega_{(n+1)}}{8 G}\tilde r_0^n\frac{\Omega  \Gamma \left(\frac{n}{2}+1\right) \,
   _2\tilde{F}_1\left(-\frac{1}{2},2;\frac{n+4}{2};\frac{A_1 (\Phi ^2_{\text{H}}-1)}{\Omega ^2-A_1}\right)}{\left(A_1-\Omega ^2\right)^2}~,\\
S=&\frac{n\Omega_{(n+1)}}{8 T G}\tilde r_0^n\frac{\, _2F_1\left(-\frac{1}{2},1;\frac{n+4}{2};\frac{A_1 (\Phi ^2_{\text{H}}-1)}{\Omega ^2-A_1}\right)}{(n+2) \left(\Omega ^2-A_1\right)}~,\\
Q=&\frac{\Omega_{(n+1)}}{16 G}\tilde r_0^n\frac{\Phi_{\text{H}}  \Gamma \left(\frac{n}{2}+1\right) \left(n \,
   _2\tilde{F}_1\left(-\frac{1}{2},1;\frac{n+4}{2};\frac{A_1 (\Phi_{\text{H}}^2-1)}{\Omega ^2-A_1}\right)\right)}{(1-\Phi_{\text{H}}^2)(\Omega ^2-A_1)}~~\\
   &+\frac{2 \,_2\tilde{F}_1\left(-\frac{1}{2},2;\frac{n+4}{2};\frac{A_1 (\Phi_{\text{H}}^2 -1)}{\Omega ^2-A_1}\right)}{(1-\Phi_{\text{H}}^2)(\Omega ^2-A_1)}~~.
\end{split}\ee


\subsubsection*{Black odd-spheres in Lifshitz background}
The thermodynamic properties of the Lifshitz black holes found in Sec.~\ref{sec:lifs} are given by
\be\begin{split}
&M=\frac{\Omega_{(n+1)}}{16\pi G}V_{(p)}r_0^n\frac{(\beta +\bold R^2)^{3/2} \bold R^{z-1} }{|\bold R^2 (z-1)+\beta  (z-2)|}~,\\
&J=\frac{\Omega_{(n+1)}}{16\pi G}V_{(p)}r_0^{n-2} n \sqrt{\beta +\bold R^2} \bold R^{z-1} \\
   &\qquad \sqrt{\frac{\left(\beta +\bold R^2\right) \left(\bold R^2 ((n+2) z+p-1)+\beta  (n (z-1)+p+2
   z-2)\right)}{\bold R^2 (n+p+2 z-1)+\beta  (n+p+2 z-2)}}~~,\\
&S=n\frac{\Omega_{(n+1)}}{16\pi T G}V_{(p)}r_0^n \sqrt{\beta +\bold R^2} \bold R^{z-1} ~,
\end{split}\ee
where we have assumed that \eqref{eq:lifcond} holds and where $\beta$ and $r_0$ are given by \eqref{eq:lifbeta} and \eqref{eq:lifr0} respectively. 


\subsubsection*{Black odd-spheres in Schwarzschild background}
In this section we collect the conserved charges for the black odd-spheres constructed in Sec.~\ref{sec:sch}. These are given by
\be\begin{split}
J=&\frac{\Omega_{(n+1)}V_{(p)}}{16\pi G}r_0^{n}\frac{R \sqrt{\tilde{\bold{R}}^{n+p}( n N\sinh ^2\alpha(n+p)+n (n+p+1)-p)+2p}}{2-\tilde{\bold{R}}^{n+p} (n+p+2)}\\
   &\sqrt{2 \left(n N \sinh ^2\alpha+n+p\right)-\tilde{\bold{R}}^{n+p} \left(2 n N \sinh ^2\alpha
   +n+p\right)}~~,\\
S=&n\frac{\Omega_{(n+1)}V_{(p)}}{16\pi T G}r_0^{n} (1-\tilde{\bold{R}}^{n+p})^{1/2} ~~,\\
\frac{Q}{\sqrt{n N} }=&\frac{\Omega_{(n+1)}V_{(p)}}{16\pi G}r_0^{n}g(\alpha)\frac{ \sqrt{\tilde{\bold {R}}^{n+p} \left(2 n N \sinh ^2\alpha+n+p\right)-2
   \left(n N \sinh ^2\alpha+n+p\right)}}{\sqrt{\left(1+N \sinh ^2\alpha\right) \left((n+p+2) \tilde{\bold {R}}^{n+p}-2\right)}}~~,
\end{split}\ee
while the mass was given in \eqref{eq:masssch}. Also the horizon radius and the chemical potential are given by
\beq
\!\!\!\!\!\!\!\!T&=&\frac{n}{4\pi r_0\cosh^N\alpha}\sqrt{\frac{n \left(1-\tilde{\bold{R}}^{n+p}\right) \left((n+p+2) \tilde{\bold{R}}^{n+p}-2\right) \left(N \sinh ^2\alpha +1\right)  }{2 n
   N \sinh ^2\alpha  \left(\tilde{\bold{R}}^{n+p}-1\right)+(n+p) \left(\tilde{\bold{R}}^{n+p}-2\right)}}~,\\
\!\!\!\!\!\!\!\!\Phi_{\text{H}}&=&\sqrt{n N} \tanh (\alpha )\sqrt{\frac{n \left(1-\tilde{\bold{R}}^{n+p}\right) \left((n+p+2) \tilde{\bold{R}}^{n+p}-2\right) \left(N \sinh ^2\alpha +1\right)  }{2 n
   N \sinh ^2\alpha  \left(\tilde{\bold{R}}^{n+p}-1\right)+(n+p) \left(\tilde{\bold{R}}^{n+p}-2\right)}}~~.
\eeq
Since we will need to compare this with the black saturn solution in App.~\ref{app:saturn} it is convenient to provide these thermodynamic properties in the uncharged limit. In this case we have
\be\begin{split} \label{eq:thermosch}
M=&\frac{\Omega_{(n+1)}V_{(p)}}{8\pi G}r_0^{n} (1-\tilde{\bold{R}}^{n+p})^{3/2} \frac{ \left(n+p+1\right)}{2-\tilde{\bold{R}}^{n+p}(n+p+2)}~~,\\
J=&\frac{\Omega_{(n+1)}V_{(p)}}{16\pi G}r_0^{n}\frac{R \sqrt{\tilde{\bold{R}}^{n+p}( n (n+p+1)-p)+2p}}{2-\tilde{\bold{R}}^{n+p} (n+p+2)}\sqrt{\left(n+p\right)(2-\tilde{\bold{R}}^{n+p})}~~,\\
S=&n\frac{\Omega_{(n+1)}V_{(p)}}{16\pi T G}r_0^{n} (1-\tilde{\bold{R}}^{n+p})^{1/2} ~~,\\
T=&\frac{n}{4\pi r_0}\sqrt{\frac{n \left(1-\tilde{\bold{R}}^{n+p}\right) \left((n+p+2) \tilde{\bold{R}}^{n+p}-2\right)   }{(n+p) \left(\tilde{\bold{R}}^{n+p}-2\right)}}~~,\\
\Omega^2 R^2=& \frac{ \left(\tilde{\bold{R}}^{n+p}-1\right) \left(\tilde{\bold{R}}^{n+p} \left(n (n+p+1)-p\right)+2 p\right)}{ \left((n+p) \left(\tilde{\bold{R}}^{n+p}-2\right)\right)}~~,\\
   \boldsymbol{\mathcal{T}}=&-\frac{\Omega_{(n+1)}V_{(p)}r_0^{n}}{16\pi G}\tilde{\bold{R}}^{n+p} \sqrt{1-\tilde{\bold{R}}^{n+p}}\frac{(n+p) \left(n+p+1\right)}{2-\tilde{\bold{R}}^{n+p}(n+p+2)}~~.
\end{split}\ee

\section{Black odd-spheres with string dipole} \label{sec:odddipole}
In this appendix we consider a related configuration to the one studied in Sec.\ref{sec:oddads} where instead of an electric charge, the black hole has a string dipole charge. The geometry is still characterised by \eqref{eq:dsodd} and the free energy \eqref{eq:freeodd}, but it must be supplemented with the polarisation vector \cite{Caldarelli:2010xz},
\beq
v^{a}\partial_a= \frac{\gamma}{R}\left(\sum_{\hat a=1}^{[(p+1)/2]}\partial_{\phi_{\hat a}}+\Omega R^2\partial_\tau\right)~~,~~ \gamma = \frac{1}{\sqrt{1-\Omega^2 R^2}}~~.
\eeq
Using the free energy \eqref{eq:freeodd} where now the dependence of $\alpha$ on $R$ is different than in the previous case, we obtain the equilibrium condition
\be
\Omega^2 R^2=(1+\bold R^2)\frac{\bold R^2 \left(n+p+1+2n N  \sinh ^2\alpha\right)+p+n N  \sinh ^2\alpha}{ \bold R^2 \left(n+p+1+2n N  \sinh ^2\alpha\right)+n+p+n N  \sinh ^2\alpha}~~.
\ee
Again, in the flat space limit, this reduces to the result obtained in \cite{Caldarelli:2010xz}. A discussion of the different limits and static cases follows similarly to the previous section. Furthermore, we obtain the tension
\be
\begin{split}
\hat{\boldsymbol{\mathcal T}}&=-\frac{V_{(p)} \Omega_{(n+1)}}{16 \pi  G}r_0^n\bold R^2 \sqrt{1+\bold R^2} \left(2nN \sinh^2\alpha+n+p+1\right)~~.
\end{split}
\ee
The thermodynamic properties can be easily obtained. In terms of the parameters $r_0,\bold R, \alpha$, these read
\be\begin{split}
M=&\frac{\Omega_{(n+1)}V_{(p)}}{16\pi G}r_0^n\left(1+\bold R^2\right)^{3/2} \left(2n N  \sinh^2\alpha+n+p+1\right)~,\\
J=&\frac{\Omega_{(n+1)}V_{(p)}}{16\pi G}r_0^{n}R\frac{ \sqrt{\left(\bold R^2 (2n N  \sinh^2\alpha+n+p+1)+p\right)}}{n }~\\
&\times \frac{\sqrt{\left(( 2n N\sinh^2\alpha+n+p+1)(1+\bold R^2 )+\bold R^2 \right)}}{n}~~,\\
S=&\frac{\Omega_{(n+1)}V_{(p)}}{4G}r_0^{n+1}  \cosh ^{ N}\alpha\sqrt{\frac{ \bold R^2\left (2n N \sinh ^2\alpha+n+p+1\right)+n N \sinh ^2\alpha+n+p}{n }}~,\\
Q=&\frac{\Omega_{(n+1)}V_{(p)}}{32\pi^2R G}r_0^{n}n\sqrt{N} \sinh \alpha\cosh \alpha~~,
\end{split}\ee
while the temperature and chemical potential are given by
\be
\begin{split}
T&= \frac{n^{3/2}}{4 \pi  r_0\cosh^{N}\alpha}   \sqrt{\frac{\bold R^2+1}{n
   \left(N \left(2 \bold R^2+1\right) \sinh ^2\alpha
   +\bold R^2+1\right)+(p+1) \bold R^2+p}}~~,\\
\Phi_{\text{H}}&=2 \pi R \sqrt{N}  \sqrt{1+\bold R^2} \tanh \alpha ~~.
\end{split}
\ee

\section{The blackfold limit of the static black saturn} \label{app:saturn}
In this section we take the blackfold limit of the black saturn solution of \cite{Elvang:2007rd} for which the spherical black hole in the centre is static and compare this with the results of Sec.~\ref{sec:sch}.

We are interred in the case for which the angular momentum of the black hole in the middle vanishes. Following the notation of \cite{Elvang:2007rd}, this means that $J^{BH}=0$ which in practice amounts to set the constant $\bar c_2=0$. In this case we have the following thermodynamic properties for the black ring surrounding the black hole
\beq \label{eq:massr}
M^{BR}=\frac{3\pi L^2}{4G}k_2~~,~~J^{BR}=\frac{\pi L^3}{G}\sqrt{\frac{k_3k_2}{2k_1}}~~,~~\Omega^{BR}=\frac{1}{L}\sqrt{\frac{k_1k_3}{2k_2}}~~,
\eeq
\beq
T^{BR}=\frac{1}{2\pi L}\sqrt{\frac{k_1(1-k_3)(k_1-k_3)}{2k_2(k_2-k_3)}}~~,~~S^{BR}=\frac{L^3\pi^2}{G}\sqrt{\frac{2k_2(k_2-k_3)^3}{k_1(k_1-k_3)(1-k_3)}}~~.
\eeq
It is also important to take a look at the mass of the black hole in the centre. This is given by
\beq \label{eq:massc}
M^{BH}=\frac{3\pi L^2}{4G}(1-k_1)~~.
\eeq
The constants $k_1,k_2,k_3$ are related to the rod intervals of the seed structure used in \cite{Elvang:2007rd} in order to obtain the solution. These are required to satisfy
\beq \label{eq:bound}
0\le k_3<k_2<k_1\le1~~,
\eeq
but only two of them are independent. The balancing condition, in order to avoid conical singularities, fixes one of them in terms of the others . This condition reads
\beq
(k_1-k_2)=\sqrt{k_1(1-k_2)(1-k_3)(k_1-k_3)}~~,
\eeq
and has two solutions but only one satisfies \eqref{eq:bound}, which reads
\beq
\!\!\!\!\!\!\!k_3=\frac{1}{2} \left(\frac{\sqrt{-k_1 (k_2-1) \left(-k_1 ((k_1-2) k_1+9) k_2+k_1
   (k_1+1)^2+4 k_2^2\right)}}{k_1 (k_2-1)}+k_1+1\right)~.
\eeq

\subsubsection*{Limits}
From the mass of the middle black hole \eqref{eq:massc}, we see that the limit in which we recover the black ring, and hence the middle black hole disappears, is the limit for which $k_1=1$. It is indeed useful to consider the reparametrization $k_1=1-\beta$ where now the black ring limit is achieved when $\beta=0$. An expansion in powers of $\beta$ is thus a weak background gravitational field expansion which can be thought of an expansion in powers of $\beta/L$ where $\beta$ is related to the middle black hole horizon and $L$ is the radius of the black ring. Here we fix the radius $L$ which formally can be set equal to one.

Furthermore, we wish to know the limit in which the black ring is ultraspinning and hence becomes effectively thin. Since we have fixed the radius $L$, this limit is achieved by sending $k_2\to0$ for which the black ring mass \eqref{eq:massr} is small and hence we have an expansion of the form $k_2/L$ where $k_2$ is proportional to the black ring horizon radius.

Therefore, to first order $\beta$ and to leading order in $k_2$ we find the following thermodynamic expressions
\beq
\Omega^{BR}=\frac{1}{2L}\left(1+\frac{1}{4}\beta\right)+\mathcal{O}(\beta^2,k_2)~,~J^{BR}=\frac{L^3\pi k_2}{2G}\left(1+\frac{5}{4}\beta\right)+\mathcal{O}(\beta^2,k_2^2)~~,
\eeq
\beq
T^{BR}=\frac{1}{2\pi L k_2}\left(1-\frac{1}{4}\beta\right)+\mathcal{O}(\beta^2, 1)~,~S^{BR}=\frac{L^3\pi^2k_2^2}{2 G}\left(1-\frac{5}{4}\beta\right)+\mathcal{O}(\beta^2,k_2^4)~.
\eeq
Furthermore, note that $M^{BR}$ does not get corrected with these expansions, neither does $M^{BH}$. To all orders in this expansion, they are given by
\beq
M^{BR}=\frac{3\pi L^2}{4G}k_2~~,~~M^{BH}=\frac{3\pi L^2}{4G}\beta~~.
\eeq 
If we set $\beta=0$ in these expressions we recover the thermodynamics of 5D ultraspinning black rings. Here we have performed a weak-field expansion in order to gain intuition regarding ultraspinning regimes of the black saturn. However, this is not necessary. To all orders in $\beta$ and to first order in $k_2$ we obtain

\beq \label{eq:obr}
\Omega^{BR}=\frac{1}{\sqrt{2}L}\sqrt{\frac{1-\beta^2}{2-\beta}}~~,~~J^{BR}=\frac{L^3\pi k_2}{\sqrt{2}G}\sqrt{\frac{1+\beta}{2-3\beta+\beta^2}}~~,
\eeq
\beq \label{eq:thermobr}
T^{BR}=\frac{1}{2L\pi k_2}\sqrt{\frac{(2-\beta)(1-\beta)^2}{2-4\beta}}~~,~~S^{BR}=\frac{\sqrt{2}L^3\pi^2k_2^2}{(1-\beta)G}\sqrt{\frac{(1-2\beta)^3}{(2-\beta)^3}}~~.
\eeq

\subsubsection*{Comparison with the blackfold construction}
We now compare this analytic solution with the construction of Sec.~\ref{sec:sch}. Using the thermodynamic charges obtained in \eqref{eq:thermosch} and taking the particular values $D=5,p=1,n=1$ we find
\beq \label{eq:balancebs}
\Omega =\frac{1}{R}\sqrt{\frac{1-\tilde{\bold{R}}^4}{2-\tilde{\bold{R}}^2}}~~,~~J=\frac{\pi R^2r_0}{2G}\frac{\sqrt{2+\tilde{\bold{R}}^2-\tilde{\bold{R}}^4}}{1-2\tilde{\bold{R}}^2}~~,
\eeq
\beq \label{eq:thermobs}
T=\frac{1}{4\pi r_0}\sqrt{\frac{1-3\tilde{\bold{R}}^2+2\tilde{\bold{R}}^4}{2-\tilde{\bold{R}}^2}}~~,~~S=\frac{2\pi^2 Rr_0^2}{G}\sqrt{\frac{2-\tilde{\bold{R}}^2}{1-2\tilde{\bold{R}}^2}}~~.
\eeq
If we now compare the angular velocity $\Omega$ given in \eqref{eq:balancebs} with the analytic solution $\Omega^{BR}$ in \eqref{eq:obr} we find that we must have
\beq\label{eq:cond}
R=\sqrt{2}L~~,~~\tilde{\bold{R}}^2=\beta~~.
\eeq
Furthermore, comparing the temperature $T$ in \eqref{eq:thermobs} with $T^{BR}$ in \eqref{eq:thermobr} we find that we must have
\beq \label{eq:cond22}
r_0=\frac{Lk_2}{\sqrt{2}}\sqrt{\frac{(1-3\tilde{\bold{R}}^2+2\tilde{\bold{R}}^4)(1-2\tilde{\bold{R}}^2)}{(2-\tilde{\bold{R}}^2)^2(1-\tilde{\bold{R}}^2)^2}}~~.
\eeq
Indeed, this identification is enough to match both the entropy and the angular momentum of the analytic solution with the blackfold approach. As for the mass, we refer to the discussion at the end of Sec.~\ref{sec:sch}, in which we have noted that the mass of the blackfold construction does not match the mass of the analytic solution $M^{BR}$. Instead we find that, in the particular case of $D=5$, we have an agreement by subtracting an appropriate factor of the tension, namely,
\beq
M^{BR}=M-\frac{1}{2}\boldsymbol{\mathcal{T}}~~.
\eeq  

\addcontentsline{toc}{section}{References}
\footnotesize
\providecommand{\href}[2]{#2}\begingroup\raggedright\endgroup

\end{document}